\documentclass[12pt, a4paper, oneside]{article}
\usepackage[top=3cm, bottom=3cm, left = 2cm, right = 2cm]{geometry} 
\usepackage{graphicx} 
\usepackage{xfrac} 
\usepackage{amsmath} 
\usepackage{amsfonts} 
\usepackage{bm}
\usepackage{comment}
\usepackage{longtable}

\usepackage{footnote}
 \usepackage{multirow}
 \usepackage{comment}

\usepackage{algorithm}
\usepackage{algorithmic}
\usepackage{mathtools} 
\usepackage{subcaption} 

\usepackage{xcolor}

\usepackage{authblk}
  \usepackage{xcolor, soul} 
\usepackage{colortbl}
\sethlcolor{lightgray}  

\usepackage{textcomp}%
\usepackage{manyfoot}%
\usepackage{booktabs}%
\usepackage{listings}%
\usepackage{comment}
\usepackage{float}

\usepackage{bm} 

\newcommand{\specialcell}[2][l]{%
\begin{tabular}[#1]{@{}l@{}}#2\end{tabular}}

\definecolor{myred}{RGB}{128,22,56} 
\usepackage[colorlinks = true,
    linkcolor = myred,
    anchorcolor = myred,
    citecolor = myred,
    filecolor = myred,
    urlcolor = myred]{hyperref} 

  \usepackage{natbib}  
\title{\vspace{-1cm}Cluster-specific ranking and variable importance for Scottish regional deprivation via vine mixtures}

\author[1]{\"{O}zge \c{S}ahin\footnote{Corresponding author: O.Sahin@tudelft.nl}}
\author[2]{Ozan Evkaya}
\author[3]{Ariane Hanebeck}

\affil[1]{\footnotesize Delft Institute of Applied Mathematics, Delft University of Technology, The Netherlands}
\affil[2]{\footnotesize School of Mathematics and Maxwell Institute for Mathematical Sciences, University of Edinburgh, UK}
\affil[3]{\footnotesize Department of Mathematics, School of Computation, Information and Technology, Technical University of Munich, Germany}
\date{}

\begin{document} 

\maketitle

\begin{abstract}
Socioeconomic deprivation is a key determinant of public health, as highlighted by the Scottish Government's Scottish Index of Multiple Deprivation (SIMD). We propose an approach for clustering Scottish zones based on multiple deprivation indicators using vine mixture models. This framework uses the flexibility of vine copulas to capture tail dependent and asymmetric relationships among the indicators. From the fitted vine mixture model, we obtain posterior probabilities for each zone's membership in clusters. This allows the construction of a cluster-driven deprivation ranking by sorting zones according to their probability of belonging to the most deprived cluster. To assess variable importance in this unsupervised learning setting, we adopt a leave-one-variable-out procedure by refitting the model without each variable and calculating the resulting change in the Bayesian information criterion. Our analysis of 21 continuous indicators across 1964 zones in Glasgow and the surrounding areas in Scotland shows that socioeconomic measures, particularly income and employment rates, are major drivers of deprivation, while certain health- and crime-related indicators appear less influential. These findings are consistent across the approach of variable importance and the analysis of the fitted vine structures of the identified clusters. \\

\textit{Keywords:} multiple deprivation, clustering, copula, vine mixture, variable importance

\end{abstract}

\section{Introduction}\label{sec:intro}
Socioeconomic deprivation is an important determinant of overall population health; therefore, many countries develop multiple deprivation indices \citep{McCartney2023}. In Scotland, the Scottish Index of Multiple Deprivation (SIMD) has served this purpose since 2004.  SIMD is one of four deprivation indices in the United Kingdom, and since its initial development, it has been heavily influenced by the early English index of multiple deprivation design in the 1990s \citep{Clelland2019}. Its version in 2020 is the Scottish Government's latest tool for identifying concentrations of deprivation in Scotland.

SIMD ranks 6976 zones, basic statistical geographies defined along natural and social boundaries, from the most deprived (rank 1) to the least (rank 6976), based on scores in seven domains (\textit{Income, Employment, Education, Health, Geographic Access to Services, Crime}, and \textit{Housing}) \citep{SIMD2020}. Each domain also has its deprivation indicators. If an area is identified as deprived, it may reflect not only low income but also fewer resources or opportunities. In practice, policymakers often focus on the zones below certain thresholds (e.g., the 5\%, 10\%, or 20\% most deprived) to prioritize interventions. Even though SIMD is widely used for tracking deprivation levels and informing policy decisions \citep{Clelland2019}, several studies question how it manages urban–rural differences, domain weights, and hidden dependencies among indicators \citep{McKendrick2011, McCartney2023}.

In that respect, clustering methods can offer an alternative for regional deprivation by grouping zones with similar profiles \citep{Senior2019}. However, to the best of our knowledge, applications of clustering to Scottish deprivation data are not common. We, therefore, propose to apply a finite mixture of vine copula-based distributions \citep{sahin2022} to cluster zones based on their main deprivation indicators in Scotland. Vine copulas allow us to detect tail dependent and asymmetric relationships among deprivation indicators that simpler models, like multivariate Gaussians, may overlook \citep{Bedford2001, Bedford2002, Aas2009}.

According to \cite{Clelland2019}, most deprived zones in Scotland are located in the Glasgow region, but its surroundings are less deprived,  making it a valuable case study for analyzing various deprivation patterns.  Therefore, we focus on Glasgow City and its surrounding areas (East/West Dunbartonshire, North/South Lanarkshire, East Renfrewshire, and Renfrewshire). We work with 21 continuous deprivation indicators spanning seven domains from socioeconomic, health, and service-access aspects across 1964 zones. Analyzing different numbers of finite mixture model components (2, 3, 4, 6, and 10), vine structures (canonical vs. regular), and clustering initialization strategies (k-means \citep{Hartigan1979} vs. Gaussian mixture models \citep{Scrucca2016}), we find that a two component R-vine mixture model with k-means initialization provides the best fit regarding the Bayesian Information Criterion (BIC) \citep{Schwarz1978}, representing the most deprived and least deprived zones.

Our vine mixture framework also provides posterior probabilities for each zone's cluster membership. We then propose a cluster-driven deprivation ranking by ordering zones according to their probability of belonging to the most deprived cluster. This probabilistic ranking reveals differences in deprivation. Moreover, policymakers may benefit from probabilistic thresholding (e.g., probabilities of belonging to the most deprived cluster exceed 0.75) rather than simply selecting a fixed quantile (e.g., the most deprived 20\% of zones). 

In addition, we propose a leave-one-variable-out (LOVO) strategy  \citep{badih2019assessing}  with BIC criterion to assess the variable importance in vine mixture models. LOVO  analysis shows which indicators are critical for clustering zones based on their deprivation indicators. In our studied zones, the main drivers of deprivation are notably \textit{Income} and \textit{Employment}. Even though the \textit{Housing} domain has the lowest weight in the latest SIMD methodology, our model suggests it is more influential than given by that. The fitted vine structures of the identified clusters also support our variable importance conclusions. 

The paper proceeds as follows. Section \ref{sec:vcmm} describes the vine mixture model, and Section \ref{sec:SIMD} outlines the current SIMD methodology from 2020. We then introduce our cluster-driven deprivation ranking and variable importance strategy in Section \ref{sec:index_vcmm}, present the data in Section \ref{sec:data}, and show empirical results in Section \ref{sec:res_cluster}. Finally, Section \ref{sec:conc} concludes the paper.

\section{Vine mixture models for clustering}\label{sec:vcmm}
Consider the scenario with $d$ deprivation indicators of $n$  zones, where a  zone (observation), $\bm{x}_i = (x_{i,1}, \ldots, x_{i,d})^\top$, is an independent realization of a $d$-dimensional random vector $\bm{X} = (X_1, \ldots, X_d)^\top$ for $i=1, \ldots, n$. It might be reasonable to expect that the dependence structure among the indicators differs based on the zones' deprivedness. For instance, in most deprived zones, a high proportion of income-deprived people may be associated with problems in the \textit{Health} domain (i.e., many emergency stays in hospitals or high mortality ratios). In contrast, these indicators might act more independently in less deprived zones. Such heterogeneity in the dependence structure suggests that a single model may fail to approximate the underlying data distribution. However, if we assume that the data arise from a finite mixture of $K$ hidden components (or groups), the joint density of $d$  indicators at  $\bm{x} = (x_1, \ldots, x_d)^\top$ can be expressed as a weighted sum of component densities. Let $\pi_k \in (0,1)$ denote the mixture weight for component $k$, such that $\sum_{k=1}^K \pi_k = 1$. Each component is parameterized by a vector $\bm{\psi}_k$, and we denote the full parameter set by $\bm{\eta} = (\bm{\eta}_1, \ldots, \bm{\eta}_K)^\top$, where $\bm{\eta}_k = (\pi_k, \bm{\psi}_k)^\top$. Then, the mixture model takes the form
\begin{equation}
g(\bm{x}; \bm{\eta}) = \sum_{k=1}^{K} \pi_k \cdot g_k(\bm{x}; \bm{\psi}_k), \label{eq:vcmm-fmm}
\end{equation}
where $g_k(\cdot)$ is the density of $k$th component for $k = 1, \ldots, K$. To obtain flexible component densities $g_k(\cdot)$, \cite{sahin2022} propose to use vine-based distributions and formulate a vine mixture model. Each $g_k(\cdot)$ in this mixture framework consists of a parametric vine copula and a set of parametric univariate marginal distributions. The vector $\bm{\psi}_k$ contains both the marginal and pair-copula parameters of the vine specification for component $k$.  

\begin{figure}[ht!]
    \centering
    \includegraphics[width=0.5\textwidth]{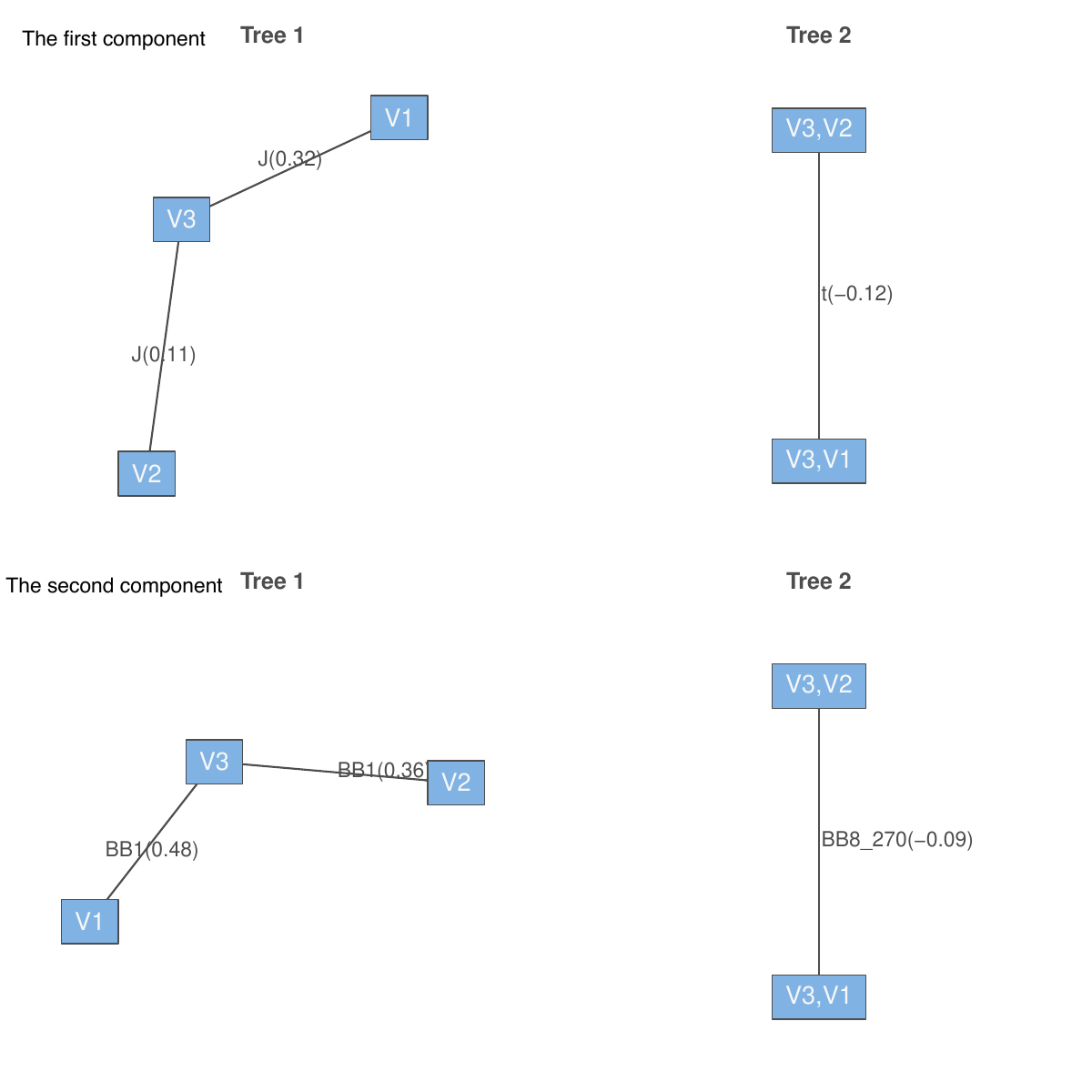}
    \caption{Vine copula model of each component, whose simulated data after coupling with margins are shown in \autoref{fig:vcmm_simdata} at the top and bottom for each component. A letter at an edge with numbers inside the parenthesis refers to its bivariate copula family with the associated Kendall's tau, where abbreviations are given in Appendix \ref{app-margin}. The variable encoding is given as follows: V1: \texttt{drive\_retail}, V2: \texttt{PT\_GP}, and V3: \texttt{PT\_retail}.}
    \label{fig:vcmm_ex_str}
\end{figure}

\begin{figure}[ht!]
    \centering
    \includegraphics[width=0.5\textwidth]{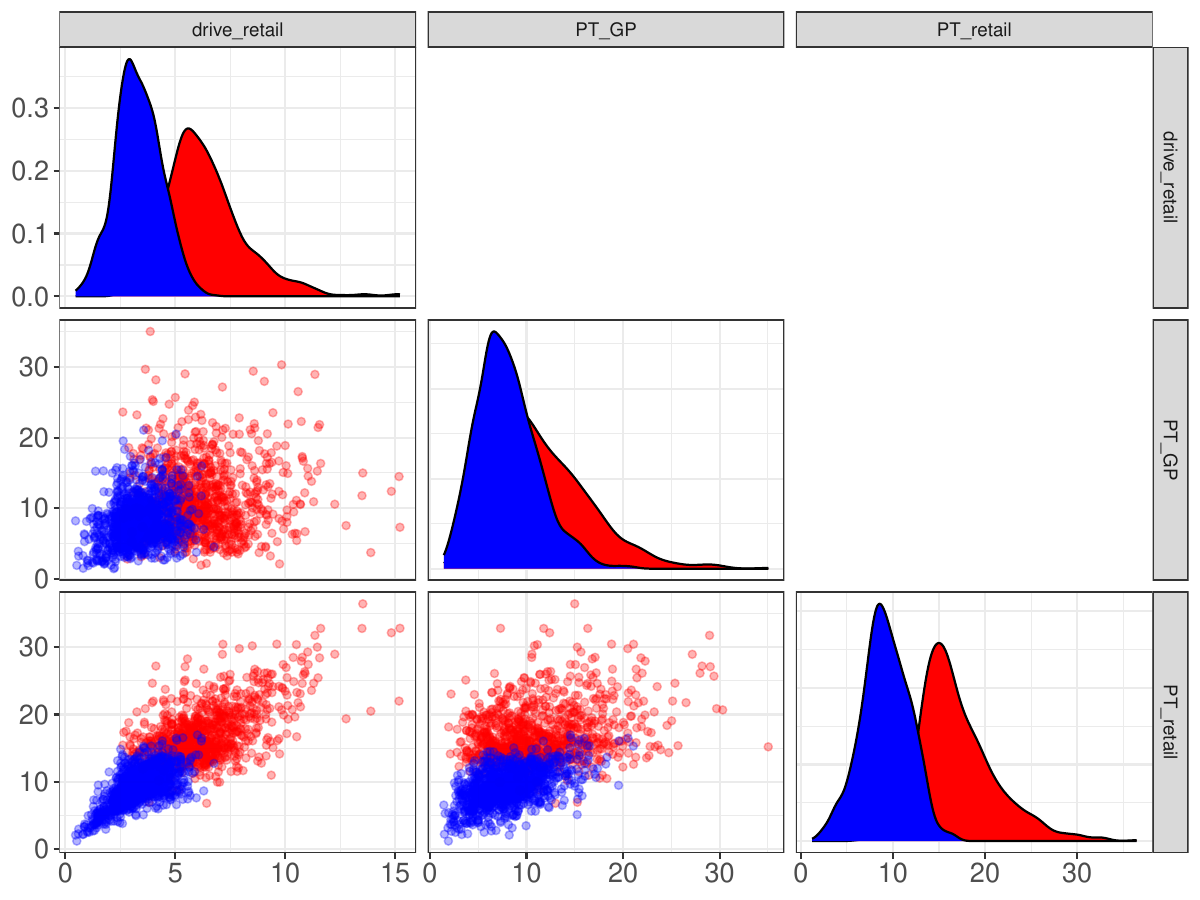}
    \caption{Pairwise scatter plot of a simulated data set (920 and 1044 observations) under the vine copula models shown in  \autoref{fig:vcmm_ex_str} and univariate marginal distributions \(X^{(1)}_1 \sim llogis(6.47, 0.16\)),
   \(X^{(1)}_2 \sim \Gamma(5.56, 0.49\)),
   \(X^{(1)}_3 \sim sstd(17.46,4.27, 4.60, 1.85\)),
 \(X^{(2)}_1 \sim snorm(3.26, 1.02, 1.15\)), \(X^{(2)}_2 \sim \Gamma(6.22, 0.77\)), \(X^{(2)}_3 \sim \mathcal{N}(9.12, 2.60\)). The blue and red points show the observations of one component and the other.}
    \label{fig:vcmm_simdata}
\end{figure}

For illustration, we simulate 1964 zones (observations) from a vine mixture model with three variables and two components. The component's simulated number of zones is 920 and 1044, respectively. Assume that three variables correspond to three indicators from the \textit{Access} domain of the SIMD data: the first one \texttt{drive\_retail}, the second one \texttt{PT\_GP}, and the third one \texttt{PT\_retail}, whose details are given in \autoref{tab:Indicators}. For simplicity, we specify the univariate margin $X^{(k)}_p$ of the $p$th indicator in the $k$th component as follows, where abbreviations are given in Appendix \ref{app-margin}. In the first component,
    \(X^{(1)}_1 \sim llogis(6.47, 0.16\));
   \(X^{(1)}_2 \sim \Gamma(5.56, 0.49\));
   \(X^{(1)}_3 \sim sstd(17.46,4.27, 4.60, 1.85\)), while in the second component, 
 \(X^{(2)}_1 \sim snorm(3.26, 1.02, 1.15\)); \(X^{(2)}_2 \sim \Gamma(6.22, 0.77\)); \(X^{(2)}_3 \sim \mathcal{N}(9.12, 2.60\)).  After simulating the data according to the specified margins and vine copula structures in \autoref{fig:vcmm_ex_str},   \autoref{fig:vcmm_simdata} shows the simulated data. We see that the vine mixture models capture heterogeneous dependence patterns in multivariate data and serve as an important modeling tool for clustering zones based on deprivation indicators.

Considering vine mixture models, the selection and estimation problems in Equation \eqref{eq:vcmm-fmm} include (i) determining the optimal number of mixture components, (ii) specifying the vine copula for the dependence structure among the $d$ indicators within each component, and (iii) selecting appropriate marginal distributions for each variable in every component. In \cite{sahin2022}, the latter two tasks are addressed using heuristic methods, while the corresponding parameter estimation is carried out via an Expectation/Conditional Maximization (ECM) algorithm \citep{Meng1993} after formulating a complete data likelihood.

More specifically, in Equation \eqref{eq:vcmm-fmm}, the true assignment of zones to clusters is unknown, and the parameter estimates depend on which cluster each zone belongs to. Therefore, one can treat the observed data as incomplete and introduce latent variables to indicate cluster membership. Specifically, for each zone $i$, we define a latent indicator vector $ 
\bm{z}_i = (z_{i,1}, \ldots, z_{i,K})^\top,
$ where $z_{i,k}$ is a binary variable with the value of 1 if zone $i$ belongs to cluster $k$ and that of 0; otherwise, with the constraint \(\sum_{k=1}^{K} z_{i,k} = 1\). 

Then, we can write the complete data likelihood as
\begin{equation}
L_c(\bm{\eta}; \bm{x}, \bm{z}) = \prod_{i=1}^{n} \prod_{k=1}^{K} \Bigl[\pi_k \, g_k(\bm{x}_i; \bm{\psi}_k)\Bigr]^{z_{i,k}}.
    \label{eq:vcmm-completell}
\end{equation}

In the E-step of the ECM algorithm, the expected values of the latent indicators $z_{i,k}$ are calculated given the current parameter estimates. Following \cite{sahin2022}, the algorithm maximizes the expected complete-data log-likelihood through three conditional maximization (CM) steps. In the first CM-step, the mixture weights $\pi_k$ are updated; in the second, the marginal parameters are optimized given the updated weights; and in the third, the pair-copula parameters are estimated conditional on the other parameter estimates. This iterative process continues until convergence, as determined by the relative change in the observed data log-likelihood between successive iterations falling below a prespecified threshold. Denote the expected value of the latent indicator $z_{i,k}$ by $r_{i,k} = \mathbb{E}[z_{i,k}],$ which corresponds to the posterior probabilities of the observation $i$ belonging to the cluster $k$.  At convergence, each observation is assigned to the cluster corresponding to the highest posterior probability, i.e., observation $i$ is assigned to the cluster $k$ for which $r_{i,k}$ is maximal.

The implementation of these procedures is provided in the \texttt{R} package \texttt{vineclust} \citep{vineclust}. For the selection of the number of components, \cite{sahin2022} propose fitting vine copula mixture models over a range of components (from 1 to 9) and then choosing the model with the smallest BIC.  In this work, we adopt the heuristic procedures from \cite{sahin2022} for specifying the vine copula and the marginal distributions in each component. To reduce the computational burden, however, we initially run the vine mixture models with a limited set of candidate component numbers, namely, 2, 4, 6, and 10. We then compare the BIC values of these models and select the two models with the smallest BICs. Next, we calculate the average component numbers of these two models and re-run the mixture model using this averaged value. If the averaged value is not an integer, we consider it to be rounded up or down based on the BIC value of the averaged ones (round towards the smaller direction). We continue this iterative procedure until convergence to a single optimal model is achieved. An example of the procedure is given in Section \ref{subsec:res_number_clust}.

\section{Scottish Index of Multiple Deprivation}\label{sec:SIMD} 
In this section, we will provide the details of the  Scottish Index of Multiple Deprivation  (SIMD) methodology and a simple numerical illustration for the general reader.  The SIMD ranks zones in Scotland from the most deprived to the least.  The way it is derived relies on a total of 32 indicators covering the seven domains with different weights: (i) \textit{Income}
($28\%$), (ii) \textit{Employment} ($28\% $), (iii) \textit{Health} ($14\%$), (iv) \textit{Education} ($14\%$), (v) \textit{Access} ($9\%$), (vi) \textit{Crime} ($5\%$), and (vii) \textit{Housing} ($2\%$) \citep{SIMD2020}.  The weights assigned to each domain originate from the Oxford University research on multiple deprivations, which are adapted based on the currency and reliability of individual indicators \citep{SIMD2020}. Later, in Section \ref{sec:index_vcmm}, we propose an alternative, cluster-driven deprivation ranking construction using deprivation indicators and assess the indicator importance for our proposed ranking.

\begin{figure}[ht!]
\centering
\includegraphics[width=1.0\linewidth]{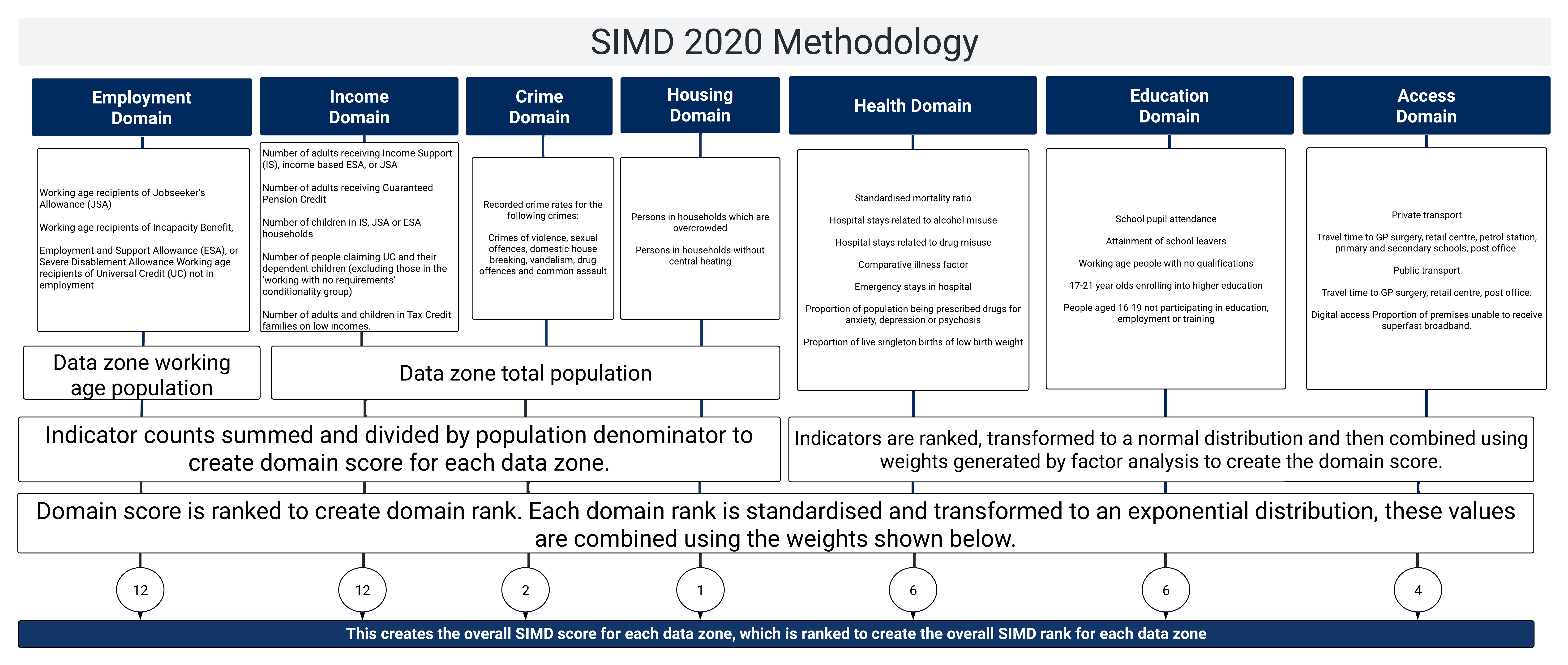}
\caption{Flowchart summarizing the Scottish Index of Multiple Deprivation (SIMD) 2020 methodology. The diagram outlines key steps: data collection from administrative sources; indicator normalization or ranking within each domain; computation of domain scores via aggregation, standardization, and, where applicable, factor-analysis weights; application of exponential transformations; assignment of domain weights; and final combination into an overall deprivation score for ranking areas. (Reproduced from the Scottish Government Technical Report 2020 via Lucidchart.)}
\label{fig:SIMD2020}
\end{figure}

To derive each domain score built on the respective indicators per zone, thereby the final deprivation index, \autoref{fig:SIMD2020} shows the current SIMD methodology. As seen in \autoref{fig:SIMD2020}, in the \textit{Income}, \textit{Employment}, and \textit{Housing} domains, the number of people corresponding to a relevant indicator  (e.g., those receiving income‐based benefits or residing in overcrowded households) is summed and then divided by the relevant population (often the 2017 mid‐year estimates from the National Records of Scotland or, in some instances, the Census) \citep{SIMD2020}. Regarding the population size, zones are assumed to be similar, each containing between 500 and 1000 households but ranging from small neighborhoods in urban settings to rural areas \citep{ScotExecutive2005}. Likewise, the \textit{Crime} domain and some \textit{Health} or \textit{Education} indicators also use these population estimates. Yet, whenever a zone has zero population in the relevant year, it is excluded from the SIMD ranking.

Even though these first‐stage calculations are straightforward for the \textit{Income}, \textit{Employment}, \textit{Crime}, and \textit{Housing}  domains since they are measured as counts per population,  \textit{Health}, \textit{Education}, and \textit{Access} domains follow a different procedure. Here, each indicator (e.g., Standardised mortality ratio, comparative illness factor) is ranked, then transformed to a standard normal distribution, and finally aggregated into a single domain score via factor analysis derived weights. Further details about the factor analysis are not provided in the technical notes; nevertheless, this extra standardization step looks essential since the indicators in these domains may involve different measurement units. For more details, including domain-specific revisions and updates, interested readers can consult the SIMD technical notes \citep{SIMD2020}.

\subsection*{SIMD ranking methodology: Example}\label{subsec:index_details}
According to the procedure summarized in \autoref{fig:SIMD2020}, for each zone in Scotland, there will be an overall SIMD score to rank from the "most deprived" to "least deprived". 

For illustration purposes, let us consider five zones (Z) belonging to the selected study region in Section \ref{sec:data}: Z1: Baillieston West, Z2: Wyndford, Z3: Blairdardie West, Z4: Ibrox, and Z5: Mosspark. For the \textit{Employment} domain, SIMD in 2020 uses two indicators: \texttt{Employment\_rate} and \texttt{Employment\_count}. They correspond to the percentage and number of employment deprived people, respectively. To derive the \texttt{Employment\_rate} per zone, SIMD uses three base quantities as seen in \autoref{fig:SIMD2020}: (i) the number of working-age recipients of Jobseeker's Allowance, (ii) the number of working-age recipients of Incapacity Benefit, Employment and Support Allowance or Severe Disablement Allowance, and (iii) the number of working-age Universal Credit claimants not in employment. To exemplify, suppose the given values of these sets are 100, 150, and 80, respectively, and the corresponding working-age population is given as 2000 for the Z1. To calculate the \texttt{Employment\_rate} indicator, thereby the \textit{Employment} domain score of Z1, we have the following steps to follow: 

\begin{enumerate}
\item[S1.] Sets of information are summed: (i) 100 + (ii) 150 + (iii) 80 = 330.
\item[S2.] Divide by the population (the zone's working-age): 
330/2000= 0.165 ($16.5\%$).
\end{enumerate}
 
Similarly, if we have the \textit{Employment} domain score of each zone (in percentage) as Z1: $16.5\%$, Z2: $43.9\%$, Z3: $3.5\%$, Z4: $30.0\%$, and Z5: $11.3\%$, we can rank these zones among each other. Similar procedures can be followed for \textit{Income}, \textit{Crime}, and \textit{Housing} domains, but using different domain-related indicators, as briefly shown in Figure  \ref{fig:SIMD2020}. 

As discussed above, for the domains of \textit{Health}, \textit{Education}, and \textit{Access}, the intermediate steps are slightly different. To illustrate, for the zone Z1 mentioned above, consider the following hypothetical numerical example in  \autoref{tab:Z1_health} using the \textit{Health} domain's weights given in \cite{SIMD2020}.  In the SIMD methodology, the indicators considered in the  \textit{Health} domain are (1) Standardised Mortality Ratio (SMR), (2) Hospital Stays Related to Alcohol Use, (3) Hospital Stays Related to Drug Use, (4) Comparative Illness Factor (CIF), (5) Emergency Stays in Hospital, (6) Proportion of Population Being Prescribed Drugs for Anxiety, Depression or Psychosis, and (7) Proportion of Live Singleton Births of Low Birth Weight in different scales. The indicators are first ranked across all zones. In \autoref{tab:Z1_health}, the third column represents the ranking of Z1 for each indicator. For example, the \texttt{SMR} indicator has the value of 92, but among Z1 to Z5, the ranking of Z1 for this indicator is assumed to be 3. Next, the rankings are transformed into standard normal z-scores as shown in the fourth column of \autoref{tab:Z1_health} among all zones. Then, the z-scores are multiplied by the weights, previously derived by factor analysis based on the workflow given in \cite{SIMD2020}. Finally, by summing up the weighted z-score, the final \textit{Health} domain score of Z1 is obtained, $0.4067416$.

\begin{table}[!h]
\centering
\caption{A hypothetical numerical illustration of the \textit{Health} domain score for Z1, where the rankings are assumed to be assigned among five zones for illustration purposes (Z1, Z2, Z3, Z4 and Z5).}
\label{tab:Z1_health}
\scalebox{0.8}{
\begin{tabular}{lrrrrr}
\toprule
Indicator & Raw value & Ranking & z-score & Weight & Weighted z-score \\
\midrule
SMR & 92 & 3 & -0.1061988 & 0.06 & -0.006371931 \\
Hospital Stays: Alcohol & 134 & 2 & -0.8495908 & 0.08 & -0.067967263 \\
Hospital Stays: Drug & 42 & 4 & 0.6371931 & 0.07 & 0.044603517 \\
CIF & 185 & 4 & 0.6371931 & 0.46 & 0.293108823 \\
Emergency Stays in Hospital & 117 & 3 & -0.1061988 & 0.19 & -0.020177781\\
Prescribed Drugs & 0.26 & 5 & 1.3805850 & 0.13 & 0.179476054 \\
Live Singleton Births & 0.07 & 1 & -1.5929827 & 0.01 & -0.015929827 \\
\hline
Sum& & & &  & 0.4067416 \\
\bottomrule
\end{tabular}
}
\end{table}

Based on the given values in Table \ref{tab:Z1_health}, one can rank the considered five zones, with the worst value (i.e., highest number, indicating more deprivation) receiving rank one and the best receiving rank five in one specific domain. Similarly, \textit{Education} and \textit{Access} domain scores can be derived using the related indicators and weights for each indicator. The above-illustrated procedure is the key to the recent SIMD rankings. When all the domain scores are derived for the considered zones, one can also create domain rankings for each.

For the completeness of the numerical example, \autoref{tab:SIMD2020_Ranks} lists the overall and domain specific SIMD rankings (as given by the SIMD 2020) of the zones shown in \autoref{tab:Z1_health}. We see that our exemplified Z1 seems to be the least deprived compared to others, where Z5 seems to be the most deprived zone among the selected five zones herein. We remark that the rankings in \autoref{tab:SIMD2020_Ranks} go beyond the five since SIMD contains 6976 zones originally. 

Further, we note that in Sections \ref{sec:data} and \ref{sec:res_cluster}, we use the raw indicator values (e.g., the second column in \autoref{tab:Z1_health}) for clustering. Alongside this, for the \textit{Employment}, \textit{Income}, \textit{Crime}, and \textit{Housing}
domains, we considered the corresponding rates, as it is mentioned in Section \ref{sec:data}.

\begin{table}[ht!]
\centering
\caption{SIMD 2020 rankings for the selected zones in Glasgow in \autoref{tab:Z1_health}. The rankings here go beyond the five since SIMD contains 6976  zones originally.}
\label{tab:SIMD2020_Ranks}
\footnotesize
\begin{tabular}{|p{1cm}|p{1cm}|p{1cm}|p{1.4cm}|p{1cm}|p{1.2cm}|p{1cm}|p{0.8cm}|p{0.8cm}|p{1.2cm}|}
\hline
 Zone & SIMD rank & Income rank & Empl. rank & Health rank & Edu. rank & Access rank & Crime rank & Hous. rank \\
\hline
Z1 & 5676 & 5885 & 5458 & 4835 & 5373 & 4167 & 1747 & 3677 \\
Z2 & 2428  & 3207 & 2334 & 2163 & 1974 & 1874  & 3873 & 2940 \\
Z3 & 79 & 88 & 84 & 62 & 507 & 4831 & 279 & 339  \\
Z4 & 1071  & 1218 & 1101 & 1148 & 1288 & 1384 & 2799 & 521 \\
Z5 & 21 & 27 & 2 & 87 & 338 & 5122& 71 & 92  \\
\hline
\end{tabular}
\end{table}

\section{Cluster-driven deprivation ranking construction and variable  importance}\label{sec:index_vcmm}

One of the main advantages of vine mixture models for clustering is that it provides posterior probabilities $r_{i,k}$ of each zone $i$ belonging to cluster $k$, for $i=1,\ldots, n$ and $k=1,\ldots, K$. Assume that we can identify a cluster $k^* \in \{1,\ldots,K\}$, which can be interpreted as the "most deprived" cluster, including the most deprived zones.
Then, the posterior probability $r_{i,k^*}$ of belonging to the most deprived cluster $k^*$ can be interpreted as the probability that zone $i$ is highly deprived.

Therefore, to construct deprivation ranks, we can sort the values of $r_{i,k^*} (i=1,\ldots,n$) from highest to lowest and assign rank 1 to the zone with the highest posterior probability to be consistent with the convention used by SIMD ranks (see Section \ref{sec:SIMD}) in our data analyses in Section \ref{sec:data}.
Ties occur if \(r_{i,k^*} = r_{j,k^*}\) for \(i \neq j\), in which case both zones $i$ and $j$ receive the same deprivation ranking.

This cluster-driven deprivation ranking measures how much more (or less) deprived one zone is relative to another based on the difference in their posterior probabilities in the most deprived cluster. Such probabilistic insights are lacking in simpler ranking systems (e.g., the current SIMD ranks), which do not report how large or small the gap between two ranks is.

One remaining question is how to identify the most deprived cluster $k^*$. Several approaches are possible. For instance, if all $d$ indicators are normalized such that lower values reflect higher deprivation, one can compute cluster-specific deprivation scores
$
s_k = \sum_{i=1}^n \sum_{p=1}^{d} r_{i,k} \,\cdot\, x^{scaled}_{i,p},
\quad \text{for} \quad k=1,\ldots,K,
$
and define $k^*$ as the cluster that minimizes $s_k$, i.e., $k^* = \arg\min_k s_k$, where $x_{i,p}$ denotes the value of indicator $p$ in zone $i$, and $x^{scaled}_{i,p}$ is its normalized value. Alternatively, one could compare boxplots of the indicator values across the clusters to identify which cluster aligns most strongly with a deprived profile. 

Next, it is important to highlight which indicators (variables) drive the clustering results, thereby the cluster-driven deprivation ranks. It is highly related to assessing variable importance in clustering. Nevertheless, there are no true labels of observations that one can analyze for variable importance in clustering (unsupervised learning). Therefore, a few alternatives include using unsupervised random forests as a clustering method \citep{breiman2001random, shi2006unsupervised}, which is not our scope for now, and the Laplacian score for variable selection \citep{he2005laplacian}.  In the latter, the main idea is to quantify how well a given variable preserves the local structure of the data manifold. However, it depends on constructing a neighborhood graph (e.g., $k$-nearest neighborhood) and often a kernel width or distance threshold. If these hyperparameters are not chosen carefully, the resulting graph may not reflect the data geometry modeled by vines.

Alternatively, \cite{badih2019assessing} propose a leave-one-variable-out (LOVO) approach and analyze how overall cluster heterogeneity changes by excluding each variable. However, in their formulation, heterogeneity is measured via covariance, which can overlook the tail and asymmetric dependence that vine copulas model. Motivated by this, we adopt a LOVO strategy tailored to BIC: for each variable excluded, we refit the vine mixture model and look at the difference in BIC relative to the full model. The indicator whose exclusion leads to the largest increase in BIC is considered the most influential. Even though this approach generalizes to distributions beyond vines and directly assesses each variable's impact on the joint likelihood, it remains computationally intensive and does not account for dependence among multiple variables. Nevertheless, it provides a practical, initial indicator ranking, which we present in Section \ref{subsec:res_newranks} for the considered data. We leave the question of how to assign relative weights (importance) of variables to future research.

\section{Data}\label{sec:data}
In this section, we provide an overview of the Scottish deprivation indicators we used for clustering in Section \ref{sec:res_cluster} and data preprocessing steps.

\subsection{Data background}
To assess the deprivation in Scotland, available domain-related indicators are described in  \autoref{tab:Indicators}, and the corresponding details on the indicators are summarized in depth in Appendix \ref{Appendix2} via \autoref{tab:Indicator_Details_app}.

\begin{figure}[ht!]
    \centering
    \includegraphics[width=0.85\linewidth]{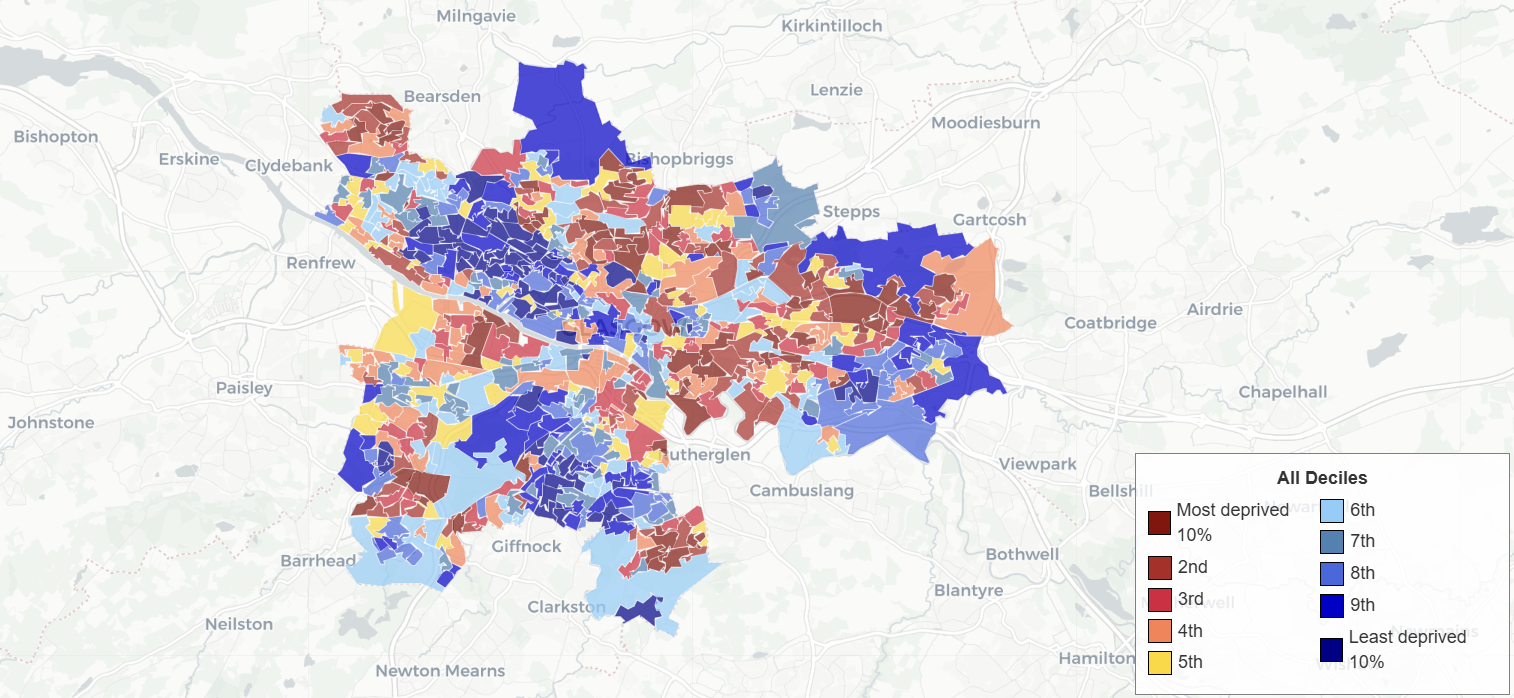}
    \caption{Glasgow City council area deprivation map, with zones shaded from dark red (most deprived) to dark blue (least deprived) and the nearby zones belonging to the surrounding council areas, North Lanarkshire, South Lanarkshire, East Renfrewshire, Renfrewshire, West Dunbartonshire, and East Dunbartonshire. Data source: Scottish Index of Multiple Deprivation (SIMD) 2020 website.}
    \label{fig:GlasgowSIMD}
\end{figure}

\begin{table}[!h]
\centering
\caption{All available indicators with their description and corresponding domain. The 21 highlighted indicators (in gray) are chosen for the analysis (Further details are given in  \autoref{tab:Indicator_Details_app}).}
\label{tab:Indicators}
\footnotesize
\scalebox{0.8}{
\begin{tabular}{ l|l|l}
\hline
 Domain & \texttt{Indicator} & Description\\\hline
 \textit{Income} & \cellcolor{lightgray}\texttt{Income\_rate} & Percentage of people who are income deprived \\
 & \texttt{Income\_count} & Number of people who are income deprived \\ \hline
 \textit{Employment} & \cellcolor{lightgray}\texttt{Employment\_rate} & Percentage of people who are employment deprived \\
 & \texttt{Employment\_count} & Number of people who are employment deprived \\\hline
 \textit{Health} & \texttt{CIF} & Comparative Illness Factor (standardised ratio) \\
 & \cellcolor{lightgray}\texttt{ALCOHOL} & Hospital stays related to alcohol use (standardised ratio) \\
 & \texttt{DRUG} & Hospital stays related to drug use (standardised ratio) \\
 & \cellcolor{lightgray}\texttt{SMR} & Standardised mortality ratio\\
 & \cellcolor{lightgray}\texttt{DEPRESS} & \specialcell{Proportion of population being prescribed drugs for anxiety,\\depression or psychosis}\\
 & \texttt{LBWT} & Proportion of live singleton births of low birth weight\\
 & \cellcolor{lightgray}\texttt{EMERG} & Emergency stays in hospital (standardised ratio)\\\hline
 \textit{Education} & \cellcolor{lightgray}\texttt{Attendance} & School pupil attendance \\
 (Education, & \cellcolor{lightgray}\texttt{Attainment} & Attainment of school leavers \\
 Skills & \cellcolor{lightgray}\texttt{no\_qualifications} & Working age people with no qualifications (standardised ratio) \\
 and Training) & \texttt{not\_participating} & \specialcell{Proportion of people aged 16-19 not participating in education,\\ employment or training}\\
 & \cellcolor{lightgray}\texttt{University} & Proportion of 17-21 year olds entering university \\\hline
 \textit{Access} & \cellcolor{lightgray}\texttt{drive\_petrol} & Average drive time to a petrol station in minutes \\
 (Geographic & \cellcolor{lightgray}\texttt{drive\_GP} & Average drive time to a GP surgery in minutes \\
 Access & \cellcolor{lightgray}\texttt{drive\_PO} & Average drive time to a post office in minutes \\
 to Services) & \cellcolor{lightgray}\texttt{drive\_primary} & Average drive time to a primary school in minutes \\
 & \cellcolor{lightgray}\texttt{drive\_retail} & Average drive time to a retail centre in minutes \\
 & \cellcolor{lightgray}\texttt{drive\_secondary} & Average drive time to a secondary school in minutes\\
 & \cellcolor{lightgray}\texttt{PT\_GP} & Public transport travel time to a GP surgery in minutes \\
 & \cellcolor{lightgray}\texttt{PT\_Post} & Public transport travel time to a post office in minutes \\
 & \cellcolor{lightgray}\texttt{PT\_retail} & Public transport travel time to a retail centre in minutes\\
 & \texttt{Broadband} & Percentage of premises without access to superfast broadband \\\hline
 \textit{Crime} & \texttt{crime\_count} & \specialcell{Number of recorded crimes of violence, sexual offences, \\ domestic housebreaking, vandalism, drugs offences,\\ and common assault} \\
 & \cellcolor{lightgray}\texttt{crime\_rate} & \specialcell{Recorded crimes of violence, sexual offences,\\ domestic housebreaking, vandalism, drugs offences,\\ and common assault per 10,000 people} \\\hline
 \textit{Housing} & \texttt{overcrowded\_count} & Number of people in households that are overcrowded \\
 & \texttt{nocentralheat\_count} & Number of people in households without central heating\\
 & \cellcolor{lightgray}\texttt{overcrowded\_rate} & Percentage of people in households that are overcrowded  \\
 & \texttt{nocentralheat\_rate} & Percentage of people in households without central heating\\\hline
\end{tabular}
}
\end{table}
While these indicators are available for all areas in Scotland, we focus on the region of Glasgow City and the surrounding council areas, specifically North Lanarkshire, South Lanarkshire, East Renfrewshire, Renfrewshire, West Dunbartonshire, and East Dunbartonshire. To illustrate, the main study region with nearby zones is visualized in \autoref{fig:GlasgowSIMD} with the levels of SIMD rankings colored differently (dark red to open blue represents the most deprived to least deprived areas, respectively). We see that the selected region contains zones with different deprivation profiles defined by the SIMD,  making the study region an attractive hot spot.

\subsection{Data preprocessing and exploratory data analysis}
For the future analysis with vine mixture models for clustering, we exclude some indicators as follows: the \texttt{rate} and \texttt{count} indicators, i.e., \texttt{Income\_}, \texttt{Employment\_}, \texttt{crime\_}, \texttt{overcrowded\_}, and \texttt{nocentralheat\_}, show high pairwise Pearson correlations of $0.9$ or higher, respectively. As the rate indicators provide more information than the count indicators by also taking the impact of the population size into account, the count indicators are omitted. Furthermore, indicators with fewer than 10\% unique values or more than 10\% zeros (as discrete or zero-inflated) in the data are removed. Using these thresholds, the indicator \texttt{CIF} is discrete while the indicators \texttt{nocentralheat\_rate}, \texttt{DRUG}, \texttt{not\_participating}, \texttt{LBWT}, and \texttt{Broadband} are zero-inflated. The histograms of these indicators are given in Appendix \ref{app-HistDiscZero}.

The above data preprocessing leads to a reduction from $32$ to $21$ indicators of $2219$ observations, which are highlighted in gray in \autoref{tab:Indicators}. Out of the $21$ indicators, \texttt{Attendance}, \texttt{Attainment}, and \texttt{crime\_rate} have $157$, $48$, and $94$ missing values, respectively. Some of them occur because data has been suppressed for reasons of disclosure control, where small numbers are involved. 
The missing values are distributed across all seven considered council areas, which exhibit between $5\%$ and $20\%$ of missing values\footnote{North Lanarkshire $5\%$, West Dunbartonshire $6\%$, Renfrewshire $12\%$, South Lanarkshire $12\%$, Glasgow City $14\%$, East Dunbartonshire $17\%$, East Renfrewshire $19\%$.}.
After removing the zones having missing values in at least one of these three indicators, the considered data set consists of $1964$ observations. Note that the indicators still cover the seven domains considered for the calculation of SIMD.

\begin{figure}[ht!]
    \centering
    \includegraphics[width=0.5\linewidth]{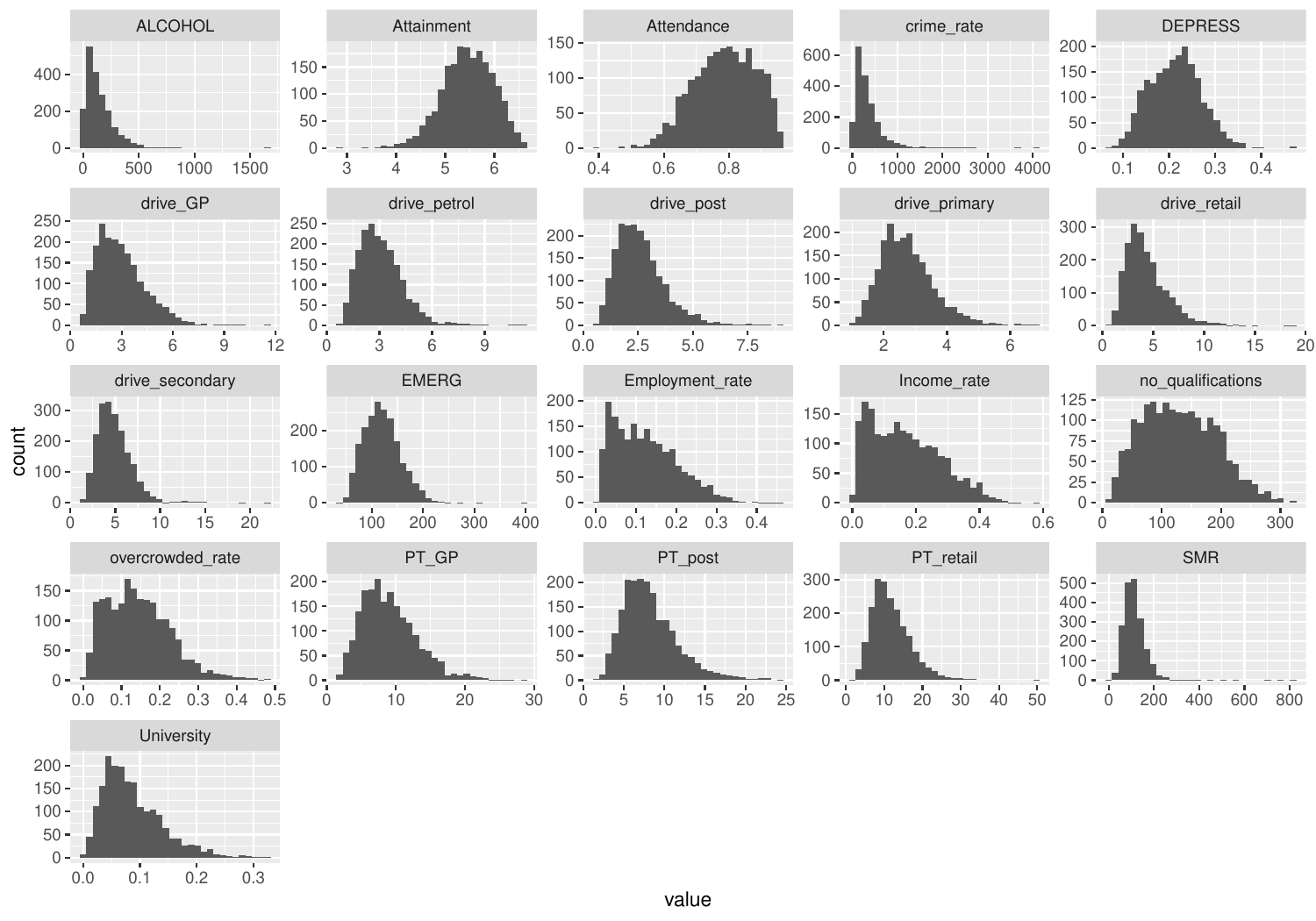}
    \caption{Histograms of the 21 considered indicators.}
    \label{fig:hist}
\end{figure}

The histograms of each considered indicator based on these $1964$ observations are given in \autoref{fig:hist}.
All examined indicators exhibit varying degrees of skewness in the data, with some being more pronounced than others. Most are positively skewed, like \texttt{ALCOHOL}, \texttt{crime\_rate}, or the indicators in the \textit{Access} domain. Others, like \texttt{Attainment} or \texttt{Attendance}, show negative skewness.
Furthermore, the indicators \texttt{ALCOHOL}, \texttt{crime\_rate}, \texttt{EMERG}, and \texttt{SMR} show some extreme values. These all occur in the council area of Glasgow city in zones like Possilpark or Drumchapel, which are known to be among the more dangerous neighborhoods of Glasgow. Moreover, we observe clear bimodality in the histogram of \texttt{overcrowded\_rate} and \texttt{Income\_rate}.

\begin{figure}[H]
    \centering
    \includegraphics[width=0.45\linewidth]{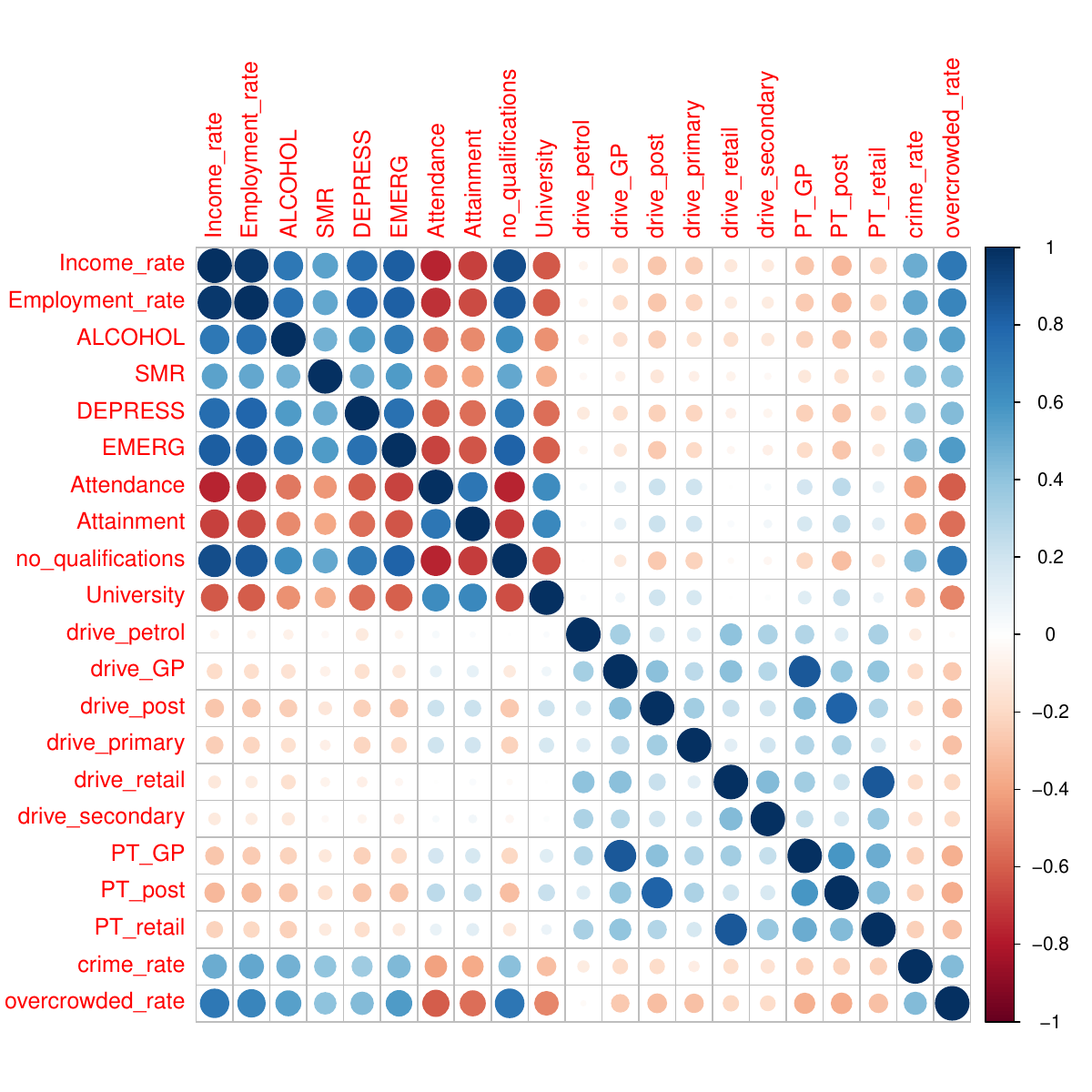}
    \caption{Correlation plot of the 21 considered indicators.}
    \label{fig:corr}
\end{figure}

\autoref{fig:corr} illustrates the pairwise Pearson correlations.
To interpret this illustration, note first that only \texttt{Attendance}, \texttt{Attainment}, and \texttt{University} are the indicators for which higher values imply lower deprivation.
This is reflected in the strongest negative correlations, which are between these three indicators and the \textit{Income}, \textit{Employment}, \textit{Health}, \textit{Housing}, and \textit{Crime} domains, as well as \texttt{no\_qualifications}.
The strongest positive correlations are observed between \textit{Income} and \textit{Employment}, as well as between these two domains and the \textit{Health}, \textit{Housing}, and \textit{Crime} domains, along with \texttt{no\_qualifications}. Furthermore, there is a strong positive correlation between the indicators in the \textit{Health} domain and between the three pairs (\texttt{PT\_GP},\texttt{drive\_GP}), (\texttt{PT\_post},\texttt{drive\_post}), and (\texttt{PT\_retail},\texttt{drive\_retail}).
Generally, the \textit{Access} domain shows only weak correlations to the other domains. Visualizing Spearman's $\rho$ leads to the same conclusions.

\section{Clustering results}\label{sec:res_cluster}
In this section, we apply vine mixture models to the deprivation data described in Section \ref{sec:data} to cluster 1964 zones with 21 continuous indicators.  Hence, we identify groups of zones with distinct deprivation characteristics by a multivariate analysis. We further examine the dependence structures and marginal distributions within the identified components. The implementation follows the \texttt{vineclust} package in \texttt{R} \citep{vineclust}. Next, we present results on the determination of the number of components, the specification of the fitted vine copula models and margins, as well as cluster-specific deprivation ranks and indicator importance ranks.

\subsection{The number of components}\label{subsec:res_number_clust}
As discussed in Section \ref{sec:vcmm}, selecting the optimal number of components in vine mixture models remains challenging. To address this, we begin our analysis with a limited set of candidate component numbers: 2, 4, 6, and 10. We consider both canonical (C-) and regular (R-) vine structures since C-vines sometimes provide lower BIC values \citep{sahin2022}. Further, we evaluate different initial partitions obtained via k-means clustering and Gaussian mixture models (GMM)  since the choice of initialization can impact the final parameter estimates.

\begin{figure}[ht!]
    \centering
    \includegraphics[width=.65\linewidth]{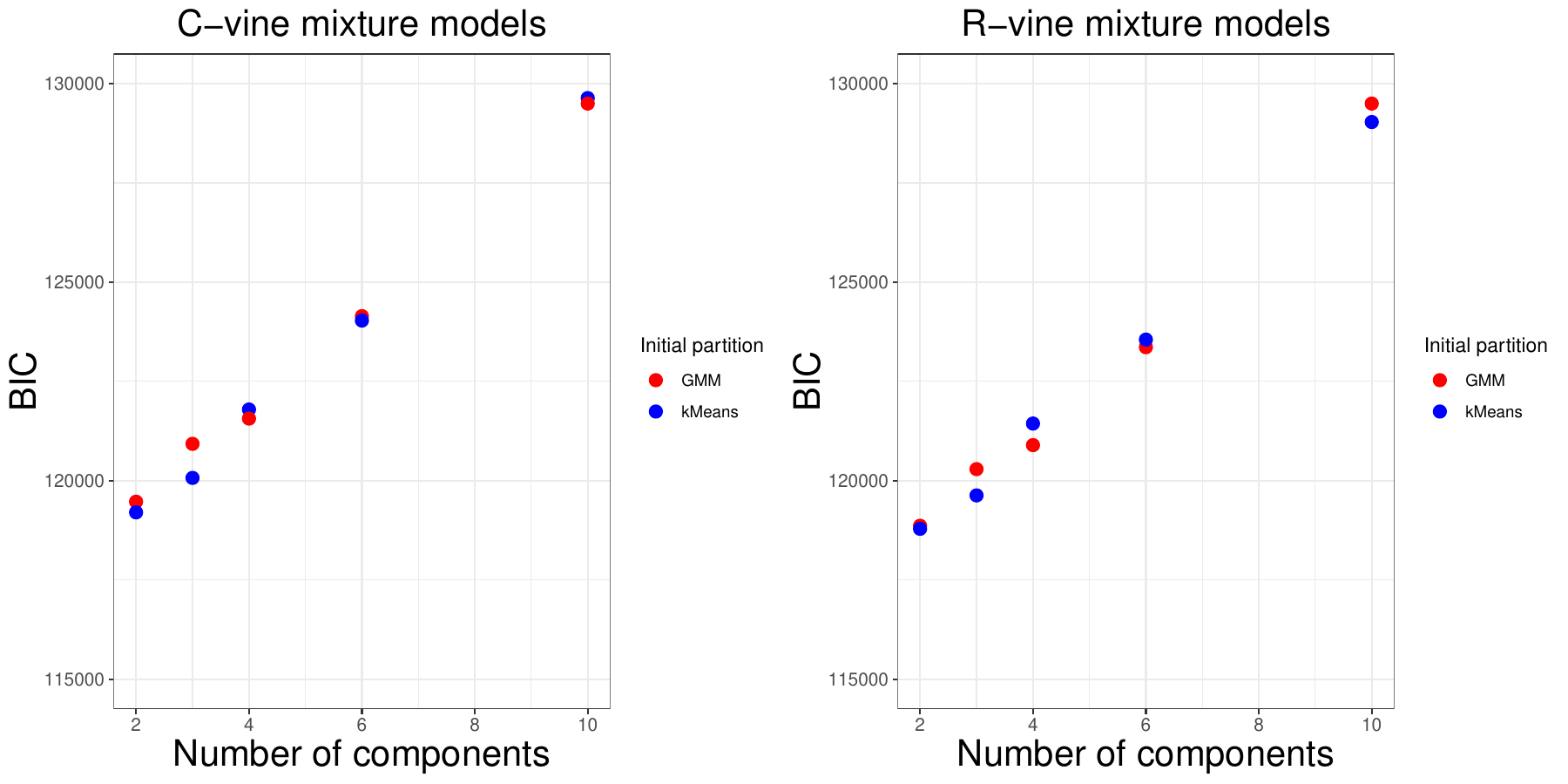}
    \caption{The BIC values of C-vine (left) and R-vine (right) mixture models with different numbers of components and initial partitions obtained via k-means clustering (kMeans) and Gaussian mixture models (GMM).}
    \label{fig:bic}
\end{figure}

\autoref{fig:bic} shows that two and four components have better BIC values than six and ten components for both structures, independent of the initial partition. The best partition, which provides the final model with the smallest BIC, changes with the number of components. Next, taking the average of two and four, we run three component C- and R-vine mixture models with two initial partitions. The resulting models have improved BIC values relative to the four component models; however, as shown in  \autoref{fig:bic}, the overall best BIC is obtained by the two component R-vine mixture model with k-means initialization. The estimated mixture weights are 0.47 and 0.53. The nearly equal weights suggest that the data have two balanced subgroups with heterogeneity in deprivation. 

We note that while our approach does not guarantee that a model with, for example, seven components would have a higher BIC than our final selection, it effectively reduces the computational burden. In our implementation, even with parallel computation for each component, one model fitting with k-means initialization takes approximately 2 hours on one core of a Linux machine (x86\_64-pc-linux-gnu) using \texttt{R} version \texttt{4.3.3}.

\subsection{Comparison of the clusters}
From the two component R-vine mixture model with k-means initialization, we have the posterior probability of each observation belonging to a cluster. Then, we assigned each observation to the cluster corresponding to the highest posterior probability and provided further cluster comparison results below.
\subsubsection{Fitted marginal distributions}
We provide the fitted marginal distribution of each indicator in each cluster in Table \ref{tab:Margins} in Appendix \ref{app:clustering_res}. We observe that the marginal distributions of 14 indicators out of 21 are modeled by different parametric families in the two clusters. Thus, the univariate distributional form of indicators is fundamentally different between clusters. For instance, for the indicator \texttt{ Employment\_rate}, Cluster 1 fits a skew-normal distribution with small location and scale parameters, while Cluster 2 models it by a Gamma distribution with parameters (7.17, 40.46). Therefore, the pattern of employment deprivation is not only shifted but also exhibits different asymmetry and variability between clusters.

In the first cluster, the heavy-tailed distributions, such as skew Student-t, are more often fitted. For instance, for \texttt{ALCOHOL} and \texttt{crime\_rate}, there is a high likelihood of extreme values in Cluster 1, whereas fitted lognormal models with a more moderate tail behavior reflect less extreme risk in Cluster 2. 

Even though \texttt{Income\_rate} is modeled by a skew-normal distribution in both clusters, the first cluster has a relatively low location (0.07) and high skewness (5) compared to the location (0.24) and skewness (1.48) in the second cluster. Therefore, there is a concentration of lower values with a long tail on the right for the first cluster. On the other hand, a shift is observed toward higher values and a more symmetric distribution in the second one.

\subsubsection{Indicators}
\begin{figure}[ht!]
    \centering
    \includegraphics[width=.65\linewidth]{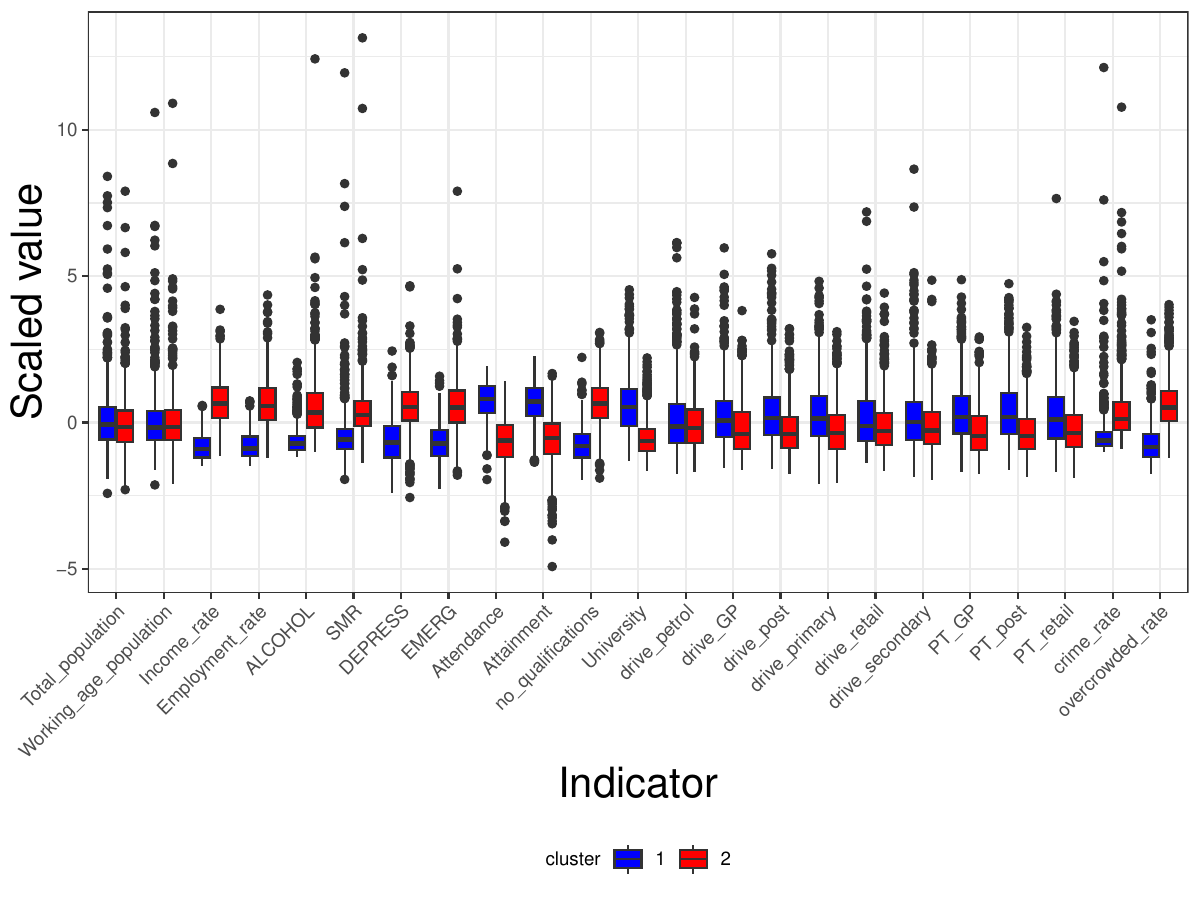}
    \caption{Boxplots of the scaled values of 21 indicators and total and working age population across clusters. We scale each variable by subtracting its estimated mean and dividing by the estimated standard deviation over all clusters.}
    \label{fig:clust_data}
\end{figure}

We observe in \autoref{tab:Margins} in Appendix \ref{app:clustering_res} that the location and scale parameters for several indicators, such as \texttt{Income\_rate}, \texttt{Attendance}, \texttt{SMR}, \texttt{drive\_post},  \texttt{drive\_retail},  \texttt{drive\_secondary}, and  \texttt{PT\_GP} differ between clusters. However, to better compare each indicator's central tendency and dispersion across clusters, we normalize each indicator by subtracting its estimated mean and dividing by its estimated standard deviation (calculated over all clusters). Then, we visualize the normalized (scaled) values of indicators in each cluster in \autoref{fig:clust_data}. We see that the first cluster, on average, has a lower crime rate and overcrowded rate, less income and employment-deprived people rate. Hence, the first cluster has better zones regarding the \textit{Income}, \textit{Employment}, \textit{Crime}, and \textit{Housing} domains of the SIMD data listed in \autoref{tab:Indicators}. The same conclusion applies to the \textit{Health} and \textit{Education} domains with the indicators \texttt{ALCOHOL}, \texttt{SMR}, \texttt{DEPRESS} and \texttt{Attendance}, \texttt{Attainment}, \texttt{no\_qualifications}, respectively, since the first cluster's average indicator values in these indicators are lower than the second one, except for \texttt{no\_qualifications} as expected (lower \texttt{no\_qualifications}, worse for a zone).

Nevertheless, the same conclusion that Cluster 1's zones do better than Cluster 2 in six domains does not apply to the \textit{Access} domain, which contains indicators about the driving or public transport time to services, such as a primary school and a general practitioner. \autoref{fig:clust_data} shows that the zones in Cluster 2 spend less time getting access to services. However, this might also be related to the possibility that zones in Cluster 1 experience more congestion.  Even though we do not consider them in clustering, for comparison, we do not observe big differences in total and working age population across clusters in \autoref{fig:clust_data}.

\begin{figure}[ht!]
    \centering
    \includegraphics[width=.5\linewidth]{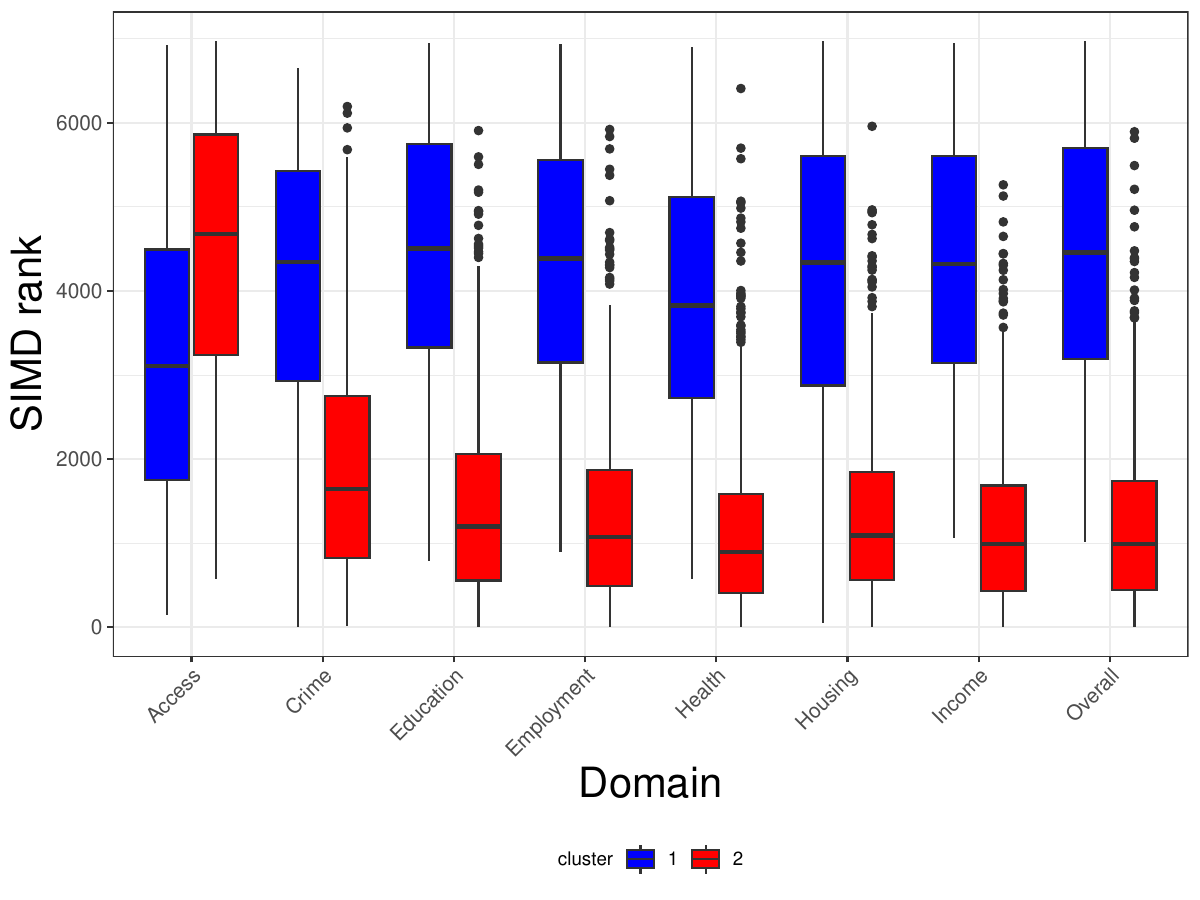}
    \caption{Comparison of the identified clusters' zones' (out of 1964 zones in Glasgow and surroundings) ranks given by SIMD for each domain and overall. The ranks go beyond 1964 since SIMD contains 6976  zones.}
    \label{fig:clust_SMID}
\end{figure}

\autoref{fig:clust_SMID} compares the identified clusters' zones' ranks given by SIMD for each domain and overall. As consistent with our analyses, the zones in the first cluster are less deprived regions regarding the SIMD. The \textit{Access} domain is the sole exception that zones in Cluster 2 are regarded as less deprived than the ones in the first one. We remark that the ranks go beyond 1964 in \autoref{fig:clust_SMID} as SIMD contains 6976  zones.

Additionally, indicator value and cluster assignment differences are observed at the council level and visualized in Figures  \ref{fig:cluster_empl_depress} and \ref{fig:cluster_empl_crime} in Appendix \ref{app:clustering_res}.

\subsubsection{Fitted vine copulas}\label{subsec:res_interp_vines}
The two fitted vine copulas corresponding to Cluster 1 and Cluster 2 clearly group the indicators of the seven considered domains in their corresponding vine tree structure. The summary of the first tree for both clusters is given in Appendix \ref{app-trees}. To illustrate this, Figure \ref{fig:illu_vine} shows a simplified illustration of their first trees, respectively, grouping indicators into their domains. The cycle between \textit{Income}, \textit{Health}, and \textit{Employment} in Cluster 2 occurs because some indicators of the \textit{Health} domain are connected to \textit{Income}, and some are connected to \textit{Employment}.

\begin{figure}[ht!]
    \centering
    \includegraphics[width=.8\linewidth]{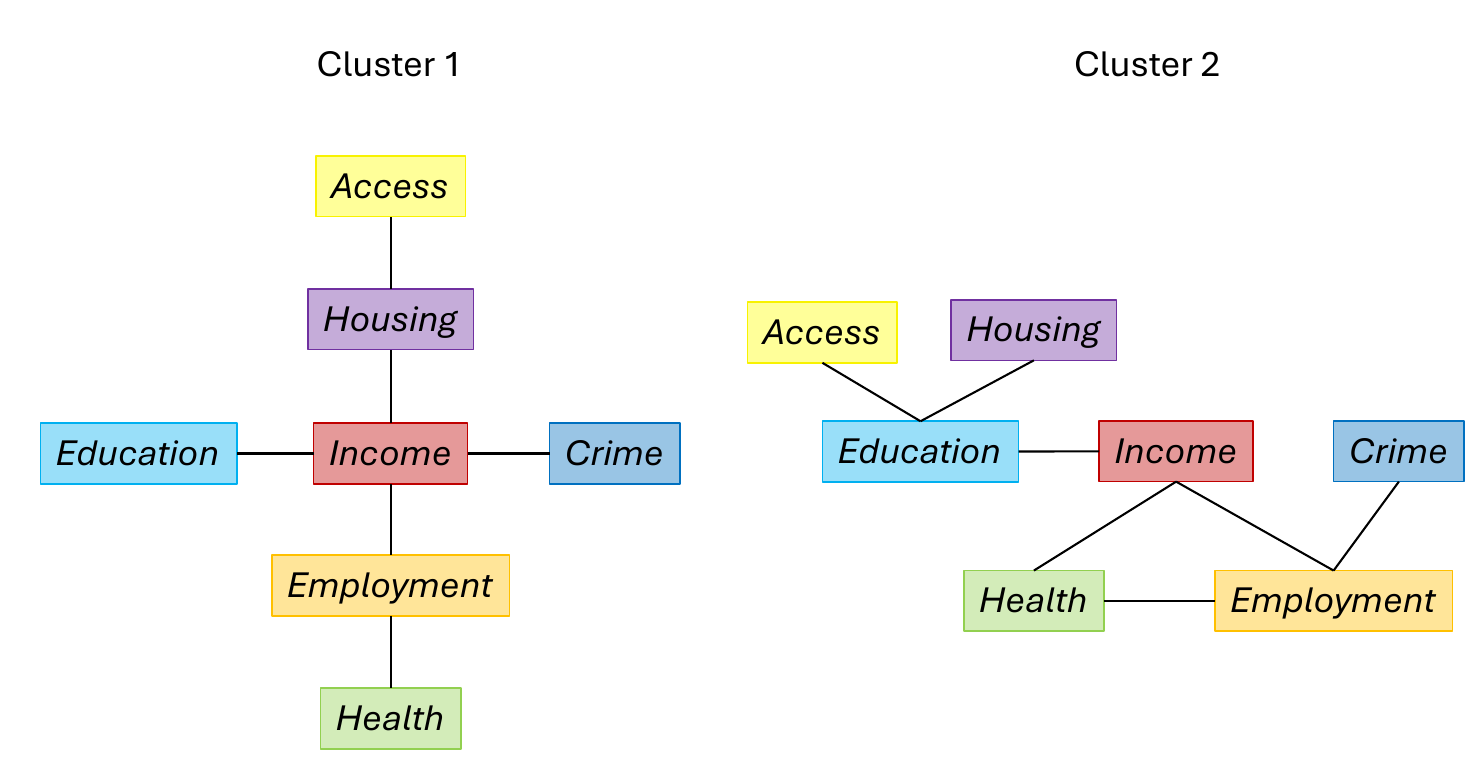}
    \caption{Simplified illustration of the first tree corresponding to the vine copula fitted for Cluster 1 (left) and Cluster 2 (right), respectively.}
    \label{fig:illu_vine}
\end{figure}

In both clusters, the \textit{Income} domain is a leverage point, being connected with four out of the remaining six domains for Cluster 1 and with three domains for Cluster 2. For both clusters, \textit{Income} and \textit{Education} are connected through a copula between \texttt{Income\_rate} and \texttt{no\_qualifications}, while \textit{Income} and \textit{Employment} are connected through \texttt{Income\_rate} and \texttt{Employment\_rate}. These copulas have an estimated positive dependence strength for both clusters, indicating that a higher percentage of people who are income-deprived and problems in the \textit{Education} and \textit{Employment} domains are highly positively related.
While the magnitude of the estimated dependence strength is similar in both clusters (Cluster 1: Gaussian copula with $\hat{\tau}=0.55$ for (\texttt{Income\_rate}, \texttt{no\_qualifications}), Frank copula with $\hat{\tau}=0.74$ for (\texttt{Income\_rate}, \texttt{Employment\_rate}) and Cluster 2: Student t copula with $\hat{\tau}=0.56$ for (\texttt{Income\_rate}, \texttt{no\_qualifications}), Gaussian copula with $\hat{\tau}=0.76$ for (\texttt{Income\_rate}, \texttt{Employment\_rate})), the t copula between \textit{Income} and \textit{Education} in Cluster 2 exhibits tail dependence, which is not given in Cluster 1.

For Cluster 1, \textit{Income} is also connected to \textit{Crime} and \textit{Housing}, which is not the case for Cluster 2.
According to Figure \ref{fig:clust_data}, Cluster 2 shows a high percentage of employment-deprived people and a high crime rate. This might lead to the association of \textit{Employment} and \textit{Crime} with a BB8 copula of $\hat{\tau}=0.33$ between \texttt{Employment\_rate} and \texttt{crime\_rate}. Cluster 2 also shows a high percentage of income-deprived people and problems in the \textit{Health} domain, which might lead to the association of \textit{Income} and the \textit{Health} domain with survival BB1 copula of $\hat{\tau}=0.47$ between \texttt{Income\_rate} and \texttt{EMERG}, exhibiting lower and upper tail dependence. For Cluster 1, these domains act more independently. Furthermore, the \textit{Education} domain plays a more important role in Cluster 2 than it does in Cluster 1, having an impact on \textit{Access} and \textit{Housing}.

\subsection{Cluster-driven deprivation ranks and variable importance}\label{subsec:res_newranks}

Following the methodology proposed in Section \ref{sec:index_vcmm}, we calculate cluster-driven deprivation ranks for all 1964 zones.  Since our optimal number of clusters is two and previous analyses (e.g., \autoref{fig:clust_data}) indicate that Cluster 2 is the most deprived, we rank the zones by their posterior probabilities of belonging to this cluster, consequently, the zone(s) with the highest posterior probability receive rank 1.  Among the 1964 zones, 52 have the same posterior probabilities with at least one other zone; thus, they are assigned the same ranking. 

\begin{figure}[ht!]
    \centering
    \includegraphics[width=.45\linewidth]{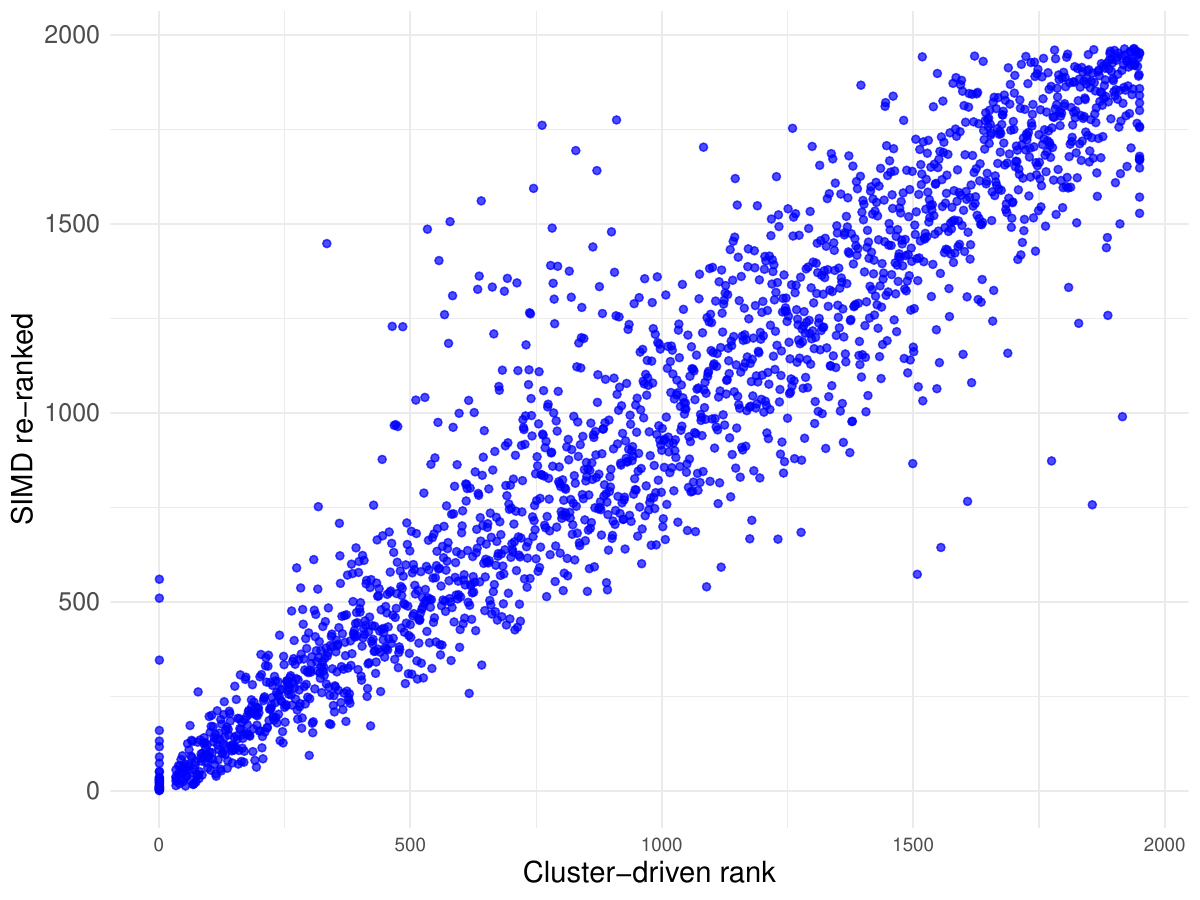}
    \caption{The comparison of vine mixture model based cluster-driven (x-axis) and SIMD given (y-axis) ranks of 1964 zones.}
    \label{fig:clust_comp_ranks}
\end{figure}

\autoref{fig:clust_comp_ranks} compares our cluster-driven ranks with the SIMD ranks, which are re-ranked to range from 1 to 1964. As seen, our ranks align quite well with the given SIMD ranks. The high alignment between the two indicates that, despite SIMD being constructed via a weighted sum of indicator values, it partially captures the underlying dependence structure modeled by our vine mixture model.

\begin{figure}[ht!]
    \centering
    \includegraphics[width=.45\linewidth]{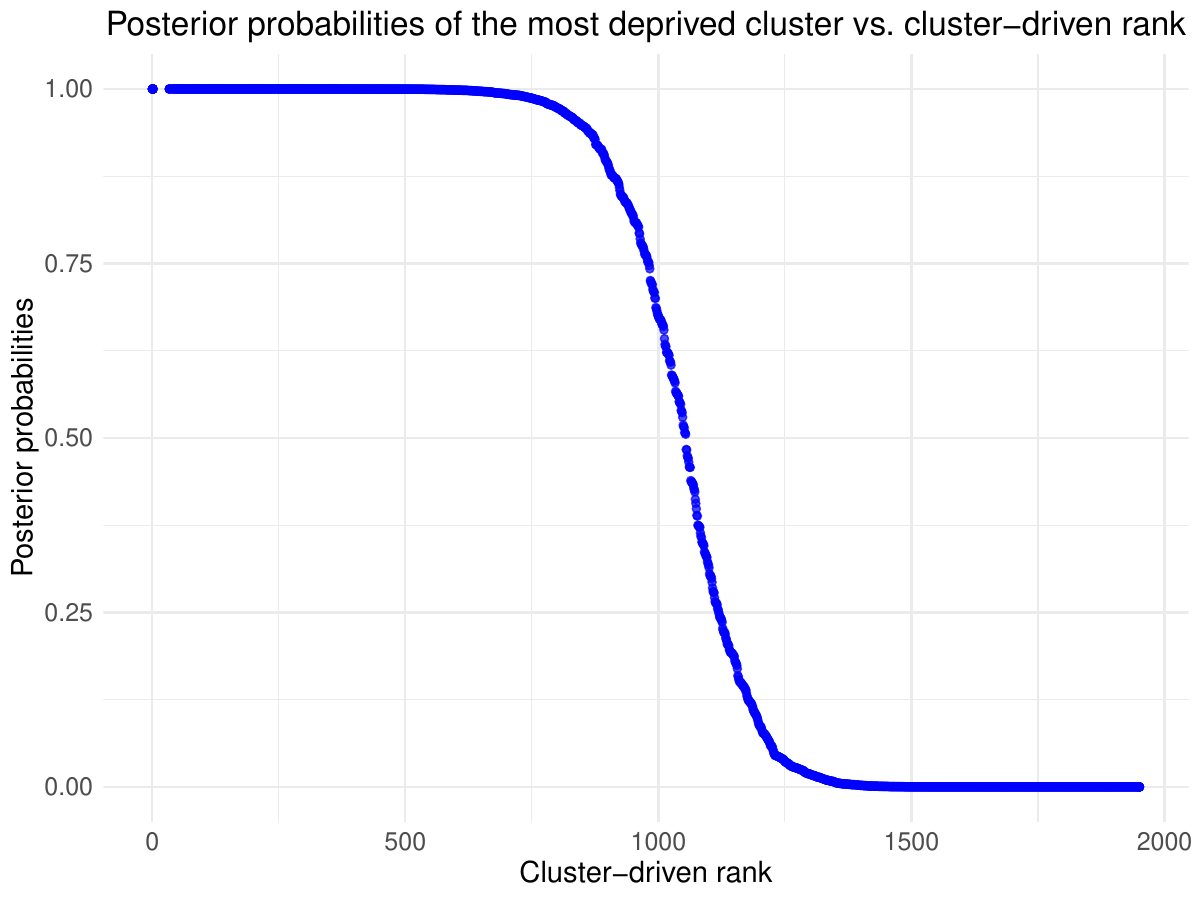}
    \caption{The posterior probabilities of each zone belonging to the most deprived cluster (y-axis) and the vine mixture model based cluster-driven (x-axis) ranks of 1964 zones.}
    \label{fig:compare_postprob_plot}
\end{figure}

\autoref{fig:compare_postprob_plot} illustrates the posterior probabilities of each zone belonging to the most deprived cluster (Cluster 2) against their cluster-driven ranks. These probabilities are nearly indistinguishable for zones ranked before approximately 750. Therefore, there are minimal differences in their deprivation risk. Similarly, from rank 1500 to around 1950, the probabilities are almost 0, implying that these zones are considered the least deprived. This insight highlights that policymakers may benefit from probabilistic thresholding rather than simply selecting a fixed quantile. For instance, one could focus further interventions on zones whose posterior probabilities of belonging to the most deprived cluster exceed 0.75.

Next, following our leave-one-variable-out (LOVO) methodology proposed in Section \ref{sec:index_vcmm}, we assess the indicator importance on the clustering via vine mixture models. Since the best model is the R-vine mixture model with two components initialized by k-means, we refit the same model setup by excluding each indicator each time, i.e., we run a 20-dimensional R-vine mixture model with two components initialized by k-means 21 times. There is no guarantee that the vine structure, initialization strategy, and number of components are the optimal setup when excluding one variable; however, we kept them the same as in the 21-dimensional full model for a fair comparison. 

\autoref{tab:lovobics} gives the LOVO strategy results of each indicator.  For each indicator, we calculate the difference between the LOVO model's BIC, which is a 20-dimensional R-vine mixture model with two components initialized by k-means without that indicator,  and the full model's BIC, which is a 21-dimensional R-vine mixture model with two components initialized by k-means. A larger increase in BIC, thereby a large absolute $\Delta$BIC, means a greater contribution of the excluded indicator to the overall model fit and, thus, its importance in deriving the cluster-driven deprivation ranks. For example, \texttt{Income\_rate} and \texttt{Employment\_rate} have large positive $\Delta$BIC values (9382.43 and 10433.19, respectively); therefore, they play a significant role in clustering. This is also consistent with the dependence structure of indicators modeled by our fitted vine copulas, whose simplified structures are shown in \autoref{fig:illu_vine}. In contrast, a negative $\Delta$BIC, as seen for \texttt{ALCOHOL} or \texttt{crime\_rate}, suggests that their exclusion may improve the mixture fit. Such improvements might result from redundancy or irrelevancy.

\begin{table}[ht!]
\centering
\caption{LOVO-BIC results for each indicator (third column) that is the BIC when each indicator is removed individually, and the corresponding change in BIC ($\Delta$BIC) (fourth column). The full model BIC (21-dimensional R-vine mixture model with two components initialized by k-means) is 118790.2. The indicator ranking based on the LOVO strategy based on $\Delta$BIC is given in the last column. }
\label{tab:lovobics}
\begin{tabular}{llrrr}
\hline
Domain & Indicator & BIC (LOVO) & \(\Delta\)BIC & Rank \\
\hline
\textit{Income}           & \texttt{Income\_rate}       & 128172.63 & 9382.43  & 2 \\
\textit{Employment}        & \texttt{Employment\_rate}   & 129223.39 & 10433.19 & 1 \\
\textit{Health}            & \texttt{ALCOHOL}            & 97454.39  & -21235.81 & 20 \\
\textit{Health}            & \texttt{SMR}                & 99896.25  & -18893.95 & 19 \\
\textit{Health}            & \texttt{DEPRESS}            & 126960.56 & 8170.36  & 3 \\
\textit{Health}            & \texttt{EMERG}              & 102244.26 & -16545.94 & 17 \\
\textit{Education}         & \texttt{Attendance}         & 124773.89 & 5983.69  & 6 \\
\textit{Education}         & \texttt{Attainment}         & 117602.20 & -1188.00 & 7 \\
\textit{Education}         & \texttt{no\_qualifications} & 100891.58 & -17898.62 & 18 \\
\textit{Education}         & \texttt{University}         & 126609.05 & 7818.85  & 4 \\
\textit{Access} & \texttt{drive\_petrol}       & 112891.90 & -5898.30  & 12 \\
\textit{Access} & \texttt{drive\_GP}           & 115757.94 & -3032.26  & 9 \\
\textit{Access} & \texttt{drive\_post}         & 115987.43 & -2802.77  & 8 \\
\textit{Access} & \texttt{drive\_primary}      & 114274.24 & -4515.96  & 10 \\
\textit{Access} & \texttt{drive\_retail}       & 114173.83 & -4616.37  & 11 \\
\textit{Access} & \texttt{drive\_secondary}    & 111294.97 & -7495.23  & 15 \\
\textit{Access} & \texttt{PT\_GP}              & 112052.26 & -6737.94  & 14 \\
\textit{Access} & \texttt{PT\_post}            & 112093.41 & -6696.79  & 13 \\
\textit{Access} & \texttt{PT\_retail}          & 110742.50 & -8047.70  & 16 \\
\textit{Crime}             & \texttt{crime\_rate}         & 93810.75  & -24979.45 & 21 \\
\textit{Housing}           & \texttt{overcrowded\_rate}    & 125907.18 & 7116.98  & 5 \\
\hline
\end{tabular}
\end{table}

As noted in Section \ref{sec:res_cluster}, SIMD assigns fixed weights to each domain (28 for \textit{Employment}, 28 for \textit{Income}, 14 for \textit{Health}, 14 for \textit{Education}, 9 for \textit{Access}, 5 for \textit{Crime}, and 2 for \textit{Housing}) when computing final deprivation ranks. Our results in \autoref{tab:lovobics} are largely consistent with these weights. However, the main difference is that the \textit{Housing} domain receives the lowest weight in SIMD, whereas our analysis indicates that the \textit{Housing} indicator ranks fifth in overall importance (out of 21). It might be since the other indicators we exclude in the \textit{Housing} domain are less important for deriving deprivation profiles. Besides, SIMD constructs its ranking by assigning different weights to indicators within each domain. For example, \texttt{EMERG} has a weight of 0.19, while \texttt{DEPRESS} is assigned 0.13 in the \textit{Health} domain. In contrast, our findings in \autoref{tab:lovobics} suggest that \texttt{DEPRESS} is more influential than \texttt{EMERG} for identifying cluster-level deprivation. On the other hand, while \texttt{CIF} has the highest weight in \textit{Health} domain by the SIMD, we removed it from our analysis due to being a discrete variable. Thus, these discrepancies and our exclusion of zero-inflated and discrete variables might explain the differences between our cluster-driven ranks and the final SIMD ranks shown in \autoref{fig:clust_comp_ranks}.

\section{Conclusion}\label{sec:conc}
In this work, we show that vine mixture models offer a powerful framework for capturing complex and often asymmetric dependence among deprivation indicators and identifying regions with similar deprivation profiles. Building on vine mixture models, we propose a probabilistic cluster‐driven deprivation ranking and a simple yet effective leave‐one‐variable‐out approach for assessing each indicator's importance in identifying the clusters (by extension, the deprivation rankings).

Focusing on Glasgow and its neighboring regions in Scotland, we show that the \textit{Income} and \textit{Employment} domains emerge as primary drivers of deprivation.  At the same time, the \textit{Health} and \textit{Education} domains presented mixed outcomes, indicating that certain indicators (e.g., \texttt{DEPRESS}, \texttt{University}) can be highly influential, whereas others (e.g., \texttt{ALCOHOL}, \texttt{no\_qualifications}) are not. Additionally, our findings suggest that indicators in \textit{Crime} and \textit{Access} tend to have lower importance, consistent with weaker dependence between these domains and core socioeconomic measures in the fitted vine structures. Hence, our results indicate that income and employment focused strategies may play a key role in coping with deprivation in Glasgow and its surroundings.

In future studies, it might be valuable to explore multi-variable leave-out strategies for variable importance in mixture models. Such an analysis might show whether groups of dependent variables jointly influence the clustering that single-variable exclusion does not capture. Further, extending the vine mixture model methodology to incorporate discrete or zero-inflated indicators, such as comparative illness factor from the health domain of the Scottish deprivation indicators, following \cite{panagiotelis2012pair} and \cite{funk2024towards} or exploring multivariate directional tail-weighted dependence measures as proposed in \cite{li2024multivariate} could deepen our understanding of how deprivation is revealed in different contexts. Additionally, creating a different cluster-based weighting through the obtained dependence structure can be worth examining further.

\newpage
\appendix

\section{SIMD 2020 Methodology Details}
\label{Appendix2}
The considered indicators in our analyses in Sections~\ref{sec:data} and \ref{sec:res_cluster} are summarized below in \autoref{tab:Indicator_Details_app}. For further details on the data source, method of construction, and key limitations, interested readers are referred to the Domains and Indicators section of the SIMD 2020 technical report \citep{SIMD2020}.

\begin{table}[ht!]
\centering
\caption{The considered 21 indicators with their description and corresponding domain.
}
\label{tab:Indicator_Details_app}
\tiny
\begin{tabular}{l|l|l}
\hline
Domain & \texttt{Indicator} & Input variables \\\hline
\textit{Income} & \cellcolor{lightgray}\texttt{Income\_rate} & \specialcell{ (i) Number of adults receiving Income Support, income-based Employment \\ and Support Allowance, or Jobseeker's Allowance (Count)}\\
& \cellcolor{lightgray} & \specialcell{(ii) Number of adults receiving Guaranteed Pension Credit (Count)} \\ 
& \cellcolor{lightgray} & \specialcell{(iii) Number of dependent children (aged 0-18) for claimants of Income \\
Support, income-based Employment and \\ Support Allowance, or Jobseeker's Allowance (Count)} \\ 
& \cellcolor{lightgray} & \specialcell{(iv) People claiming Universal Credit and their dependent children \\ (excluding those in the 'working with no requirements'
conditionality group)  (Count)} \\ 
& \cellcolor{lightgray} & \specialcell{(v) Number of adults and children in Tax Credit families on low
incomes (Count)} \\ 
\hline
\textit{Employment} & \cellcolor{lightgray}\texttt{Employment\_rate} & (i) Working age recipients of Jobseeker's Allowance (Count) \\
& \cellcolor{lightgray} & \specialcell{(ii) Working age recipients of Incapacity Benefit, Employment and \\
Support Allowance or Severe Disablement Allowance(Count)} \\
& \cellcolor{lightgray} & \specialcell{(iii) Working age Universal Credit claimants not in employment (Count)} \\
\hline
\textit{Health} & \cellcolor{lightgray}\texttt{ALCOHOL} & \specialcell{Indirectly age-sex standardised ratio of observed to
expected stays \\ in acute NHS hospitals in Scotland with a
diagnosis of alcohol-related conditions \\ (based on any of six possible diagnoses), both sexes, all ages (Indirectly standardised ratio)} \\
& \cellcolor{lightgray}\texttt{SMR} & \specialcell{ Indirectly age-sex standardised ratio for deaths of all ages registered \\ from all causes. Data standardised by 5-year age band and sex \\ (Indirectly standardised ratio)} \\
& \cellcolor{lightgray}\texttt{DEPRESS} & \specialcell{ Proportion of patients being prescribed anxiolytic,
antipsychotic \\ or antidepressant drugs. Derived from paid
prescriptions data at patient level (Proportion)}\\
 & \cellcolor{lightgray}\texttt{EMERG} & \specialcell{ Indirectly age-sex standardised ratio of observed to
expected emergency \\ stays in acute NHS hospitals in
Scotland, both sexes and all ages \\ (Indirectly standardised ratio)}\\
 \hline
 \textit{Education} & \cellcolor{lightgray}\texttt{Attendance}
 & \specialcell{Percentage of pupils who attend school 90\% or more of the time for each  zone\\ in Scotland; includes pupils who attend publicly funded primary, secondary and special \\ schools (Percentage of pupils with attendance of 90\% or above
over a two year period)}\\
  & \cellcolor{lightgray}\texttt{Attainment} & \specialcell{Score based on school leavers' highest level of the qualification for \\ pupils who attend publicly funded secondary schools (Average score-three year average)} \\
  & \cellcolor{lightgray}\texttt{no\_qualifications} & \specialcell{The indicator shows the percentage of working age adults (aged 25-64) that \\ were recorded as having no qualifications in the 2011 Census (Indirectly standardised ratio.)} \\
 & \cellcolor{lightgray}\texttt{University} & \specialcell{The indicator reflects the enrolment rate based on the number of 17-21 entrants to first \\degree courses domiciled in each zone before the start of their course, compared \\to the total number of 17-21 year olds resident in that zone over \\the same period (Rate-over three years)} \\
 \hline
 \textit{Access} & \cellcolor{lightgray}\texttt{drive\_petrol} & \specialcell{Drive time sub-domain (weight in access domain: 60\%) \\ Population weighted average drive time taken to reach petrol station (in minutes)} \\
  & \cellcolor{lightgray}\texttt{drive\_GP} & \specialcell{Population weighted average drive time taken to reach GP (in minutes)} \\
  & \cellcolor{lightgray}\texttt{drive\_PO} & \specialcell{Population weighted average drive time taken to reach post office (in minutes)} \\
 & \cellcolor{lightgray}\texttt{drive\_primary} & \specialcell{Population weighted average drive time taken to reach primary school (in minutes)} \\
 & \cellcolor{lightgray}\texttt{drive\_retail} & \specialcell{Population weighted average drive time taken to reach retail centre (in minutes)} \\
 & \cellcolor{lightgray}\texttt{drive\_secondary} & \specialcell{Population weighted average drive time taken to reach secondary school (in minutes)} \\
 & \cellcolor{lightgray}\texttt{PT\_GP} & \specialcell{Public transport sub-domain (weight in access domain: 30\%) \\ Population weighted average travel time to GP (in minutes)} \\
 & \cellcolor{lightgray}\texttt{PT\_Post} & \specialcell{Population weighted average travel time to post office (in minutes)} \\
 & \cellcolor{lightgray}\texttt{PT\_retail} & \specialcell{Population weighted average travel time to retail centre (in minutes)}  \\
 \hline
 \textit{Crime} & \cellcolor{lightgray}\texttt{crime\_rate} & 
 \specialcell{Recorded crime rate of selected crimes of violence, sexual offences, domestic housebreaking, \\vandalism, drug offences and common assault. The overall indicator is a sum of each SIMD \\crime per 10,000 population (Rate per 10,000 people)} \\\hline
 \textit{Housing} & \cellcolor{lightgray}\texttt{overcrowded\_rate} & \specialcell{This indicator provides a measure of material living standards and gives the proportion of \\ household population that lives in overcrowded housing based on the occupancy rating. \\Overcrowding is defined to mean households with an occupancy rating of -1 or less.
\\ This means that there is at least one room too few in the household \\(Percentage of household population)} \\
 \hline
\end{tabular}
\end{table}

\section{Abbreviation for univariate marginal distributions and bivariate copulas}\label{app-margin}
For bivariate univariate marginal distributions, we use the following abbreviations: $cauchy(a,b)$: Cauchy distribution with location parameter $a$ and scale parameter $b$,
$snorm(a,b,c)$: skew normal distribution with location parameter $a$, scale parameter $b$, and skewness parameter $c$,
$lnorm(\mu,\sigma)$: log-normal distribution with mean/standard deviation parameters $\mu$/$\sigma$ on the logarithmic scale, 
$exp(\lambda)$: exponential distribution with rate parameter $\lambda$, 
$llogis(\alpha, \beta)$: log-logistic distribution with shape parameter $\alpha$ and scale parameter $\beta$, 
$logis(l, s)$: logistic distribution with location parameter $l$ and scale parameter $s$,  
$\Gamma(\alpha,\beta)$: gamma distribution with shape parameter $\alpha$ and rate parameter $\beta$,  
$\mathcal{N}(\mu,\sigma)$: normal distribution with mean parameter $\mu$ and standard deviation parameter $\sigma$,
$t(\mu,\sigma, \nu)$: Student's t distribution with location parameter $\mu$, scale parameter $\sigma$, and shape parameter $\nu$,
$sstd(a,b,c,d)$: skew Student’s t distribution with location parameter $a$, scale parameter $b$, shape parameter $c$, and skewness parameter $d$.

For bivariate copulas, we use the following abbreviations: N: Gaussian, SG: Survival Gumbel, F: Frank, BB1: BB1, SBB8: Survival BB8,  t: t, SC: Survival Clayton, R2C: Rotated Clayton 270 degrees, R9G: Rotated Gumbel 90 degrees, R2G: Rotated Gumbel 270 degrees,  R9J: Rotated Joe 90 degrees,  R2BB1: Rotated BB1 270 degrees copula.

\section{Histograms of discrete and zero-inflated indicators}\label{app-HistDiscZero}

\begin{figure}[H]
    \centering
    \includegraphics[width=.75\linewidth]{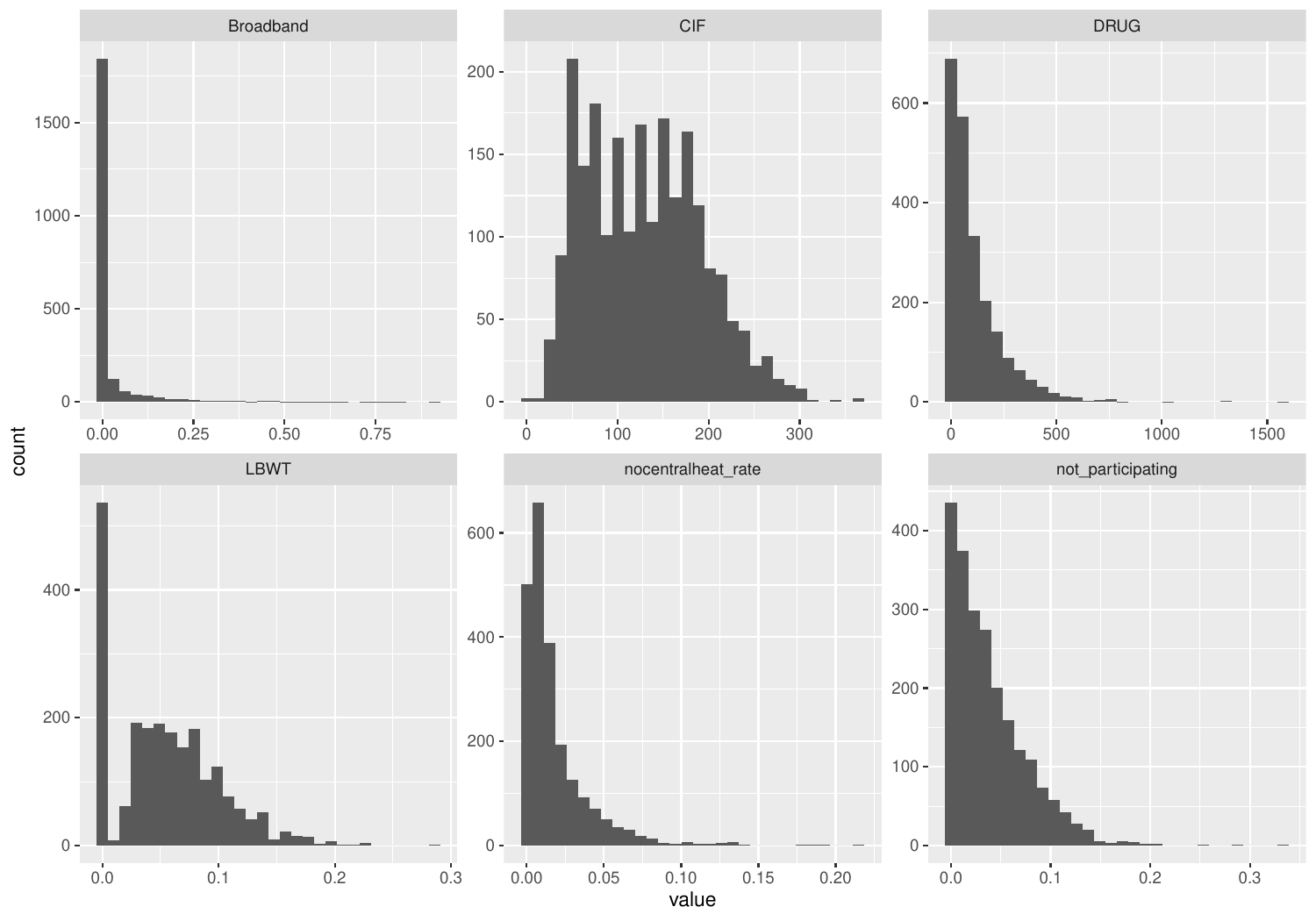}
    \caption{Histograms of the discrete and zero-inflated indicators, which are omitted for the data analysis, based on $2219$ observations. The indicator \texttt{CIF} only has 3\% of unique values, while the indicators \texttt{Broadband}, \texttt{DRUG}, \texttt{LBWT}, \texttt{nocentralheat\_rate}, and \texttt{not\_participating} show 70\%, 17\%, 24\%, and 10\%, and 19\% of zeros in the data, respectively.}
    \label{fig:histDiscZero}
\end{figure}

\clearpage
\section{Clustering results}\label{app:clustering_res}

\begin{table}[!h]
\centering
\caption{The fitted marginal distributions for each indicator in each cluster.}
\label{tab:Margins}
\begin{tabular}{llll}
  \hline
  Indicator & Cluster 1 & Cluster 2 \\ 
  \hline
  \texttt{Income\_rate} &  $snorm(0.07, 0.05, 5)$ & $snorm(0.24, 0.08, 1.48)$  \\ 
  \texttt{Employment\_rate} & $snorm(0.06, 0.03, 2.98)$ & $\Gamma(7.17, 40.46)$ \\ 
  \texttt{ALCOHOL} & $sstd(66.24, 52.08, 6.22, 3.16)$ & $lnorm(5.2, 0.57)$ \\ 
  \texttt{SMR} & $llogis(4.68, 0.01)$ & $llogis(6.12, 0.01)$ \\ 
  \texttt{DEPRESS} & $\Gamma(18.36, 103.36)$ & $logis(0.24, 0.03)$ \\ 
  \texttt{EMERG} & $\Gamma(18.47, 0.19)$ & $llogis(8.27, 0.01)$ \\ 
  \texttt{Attendance} & $snorm(0.86, 0.06, 0.56)$ & $snorm(0.73, 0.07, 0.87)$  \\ 
  \texttt{Attainment} & $snorm(5.82, 0.36, 0.8)$ & $sstd(5.15, 0.43, 13.56, 0.83)$ \\ 
  \texttt{no\_qualifications} & $\Gamma(4.88, 0.06)$ & $\mathcal{N}(171.51, 48.18)$ \\ 
  \texttt{University} & $\Gamma(5.17, 43.08)$ & $snorm(0.06, 0.03, 1.81)$ \\ 
  \texttt{drive\_petrol} & $lnorm(1.07, 0.44)$ & $\Gamma(7.11, 2.38)$ \\ 
  \texttt{drive\_GP} & $lnorm(1.11, 0.44)$ & $snorm(2.68, 1.3, 2.8)$ \\ 
  \texttt{drive\_post} & $lnorm(1.02, 0.39)$ & $lnorm(0.76, 0.4)$ \\ 
  \texttt{drive\_primary} & $lnorm(1.05, 0.3)$ & $\Gamma(12.21, 4.84)$ \\ 
  \texttt{drive\_retail} & $lnorm(1.43, 0.46)$ & $lnorm(1.3, 0.45)$ \\ 
  \texttt{drive\_secondary} & $lnorm(1.57, 0.4)$ & $lnorm(1.47, 0.35)$ \\ 
  \texttt{PT\_GP} & $\Gamma(6.54, 0.62)$ & $\Gamma(5.37, 0.68)$ \\ 
  \texttt{PT\_post} & $lnorm(2.2, 0.38)$ & $\Gamma(7.49, 1.06)$ \\ 
  \texttt{PT\_retail} & $lnorm(2.5, 0.41)$ & $\Gamma(6.33, 0.6)$ \\ 
  \texttt{crime\_rate} & $sstd(189.95, 219.95, 2.6, 4.43)$ & $lnorm(5.94, 0.61)$ \\ 
  \texttt{overcrowded\_rate} & $sstd(0.08, 0.05, 11.5, 2.34)$ & $\Gamma(8.37, 42.47)$ \\ 
 \hline
\end{tabular}
\end{table}

 We observe that most zones in Glasgow City are assigned to Cluster 2 (most deprived), whereas East Dunbartonshire and East Renfrewshire have the zones assigned to Cluster 1. While a higher employment deprived rate is positively related to the population being prescribed drugs for depression in all council areas, such a relation is not clearly seen for the  \texttt{Employment\_rate} and \texttt{crime\_rate}. Some zones experience higher crime rates despite similar employment levels, open to further investigations by policymakers. 
 
\begin{figure}[H]
\centering
\includegraphics[width=0.75\textwidth]{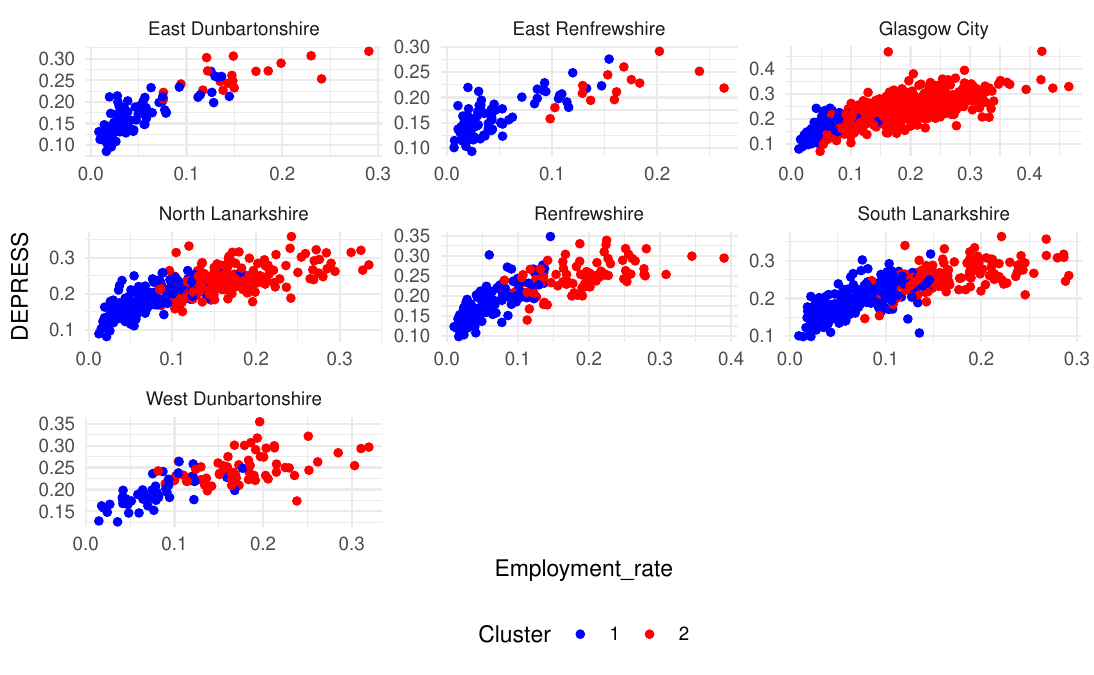}
\caption{Scatter plots of the identified clusters' zones'  \texttt{Employment\_rate}  and \texttt{DEPRESS} across different council areas.}
\label{fig:cluster_empl_depress}
\end{figure}

\begin{figure}[H]
\centering
\includegraphics[width=0.75\textwidth]{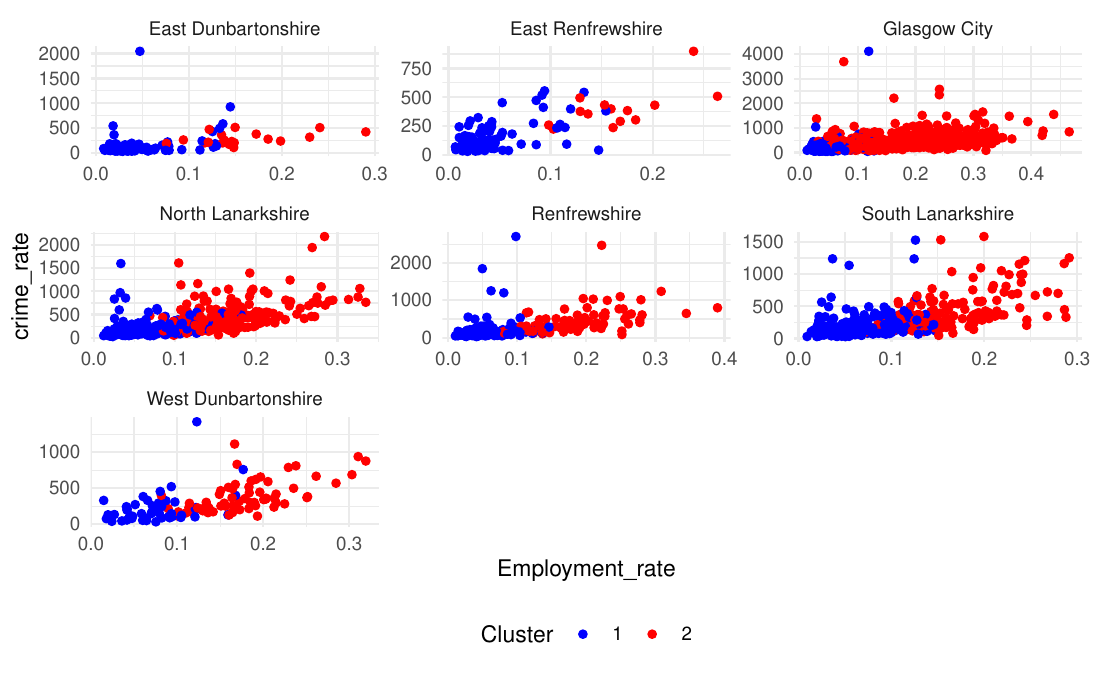}
\caption{Scatter plots of the identified clusters'  zones' \texttt{Employment\_rate}  and \texttt{crime\_rate} across different council areas.}
\label{fig:cluster_empl_crime}
\end{figure}

\section{Fitted vine copulas of clusters
}\label{app-trees}

We give the summary of the first tree of the vine copula fitted for Cluster 1 and Cluster 2, respectively. Here, 1: \texttt{Income\_rate},   2: \texttt{Employment\_rate},   3: \texttt{ALCOHOL},   4: \texttt{SMR},   5: \texttt{DEPRESS},   6: \texttt{EMERG}, 7: \texttt{Attendance},   8: \texttt{Attainment},   9: \texttt{no\_qualifications},   10: \texttt{University},  11: \texttt{drive\_petrol}, 12: \texttt{drive\_GP},   13: \texttt{drive\_post},   14: \texttt{drive\_primary},   15: \texttt{drive\_retail},   16: \texttt{drive\_secondary}, 17: \texttt{PT\_GP},   18: \texttt{PT\_post},   19: \texttt{PT\_retail},   20: \texttt{crime\_rate},   21: \texttt{overcrowded\_rate}.
\begin{table}[ht]
\centering
\begin{tabular}{rrlrlrrrrr}
  \hline
 & tree & edge & family & cop & par & par2 & tau & utd & ltd \\ 
  \hline
1 &   1 & 19,17 & 2.00 & t & 0.50 & 5.07 & 0.33 & 0.20 & 0.20 \\ 
  2 &   1 & 19,15 & 1.00 & N & 0.87 & 0.00 & 0.67 & 0.00 & 0.00 \\ 
  3 &   1 & 17,18 & 2.00 & t & 0.61 & 4.34 & 0.41 & 0.30 & 0.30 \\ 
  4 &   1 & 18,13 & 1.00 & N & 0.81 & 0.00 & 0.60 & 0.00 & 0.00 \\ 
  5 &   1 & 17,12 & 17.00 & SBB1 & 0.21 & 2.51 & 0.64 & 0.27 & 0.68 \\ 
  6 &   1 & 15,16 & 6.00 & J & 1.62 & 0.00 & 0.26 & 0.47 & 0.00 \\ 
  7 &   1 & 15,11 & 2.00 & t & 0.40 & 3.87 & 0.26 & 0.21 & 0.21 \\ 
  8 &   1 & 13,14 & 4.00 & G & 1.27 & 0.00 & 0.21 & 0.27 & 0.00 \\ 
  9 &   1 & 14,21 & 5.00 & F & -1.74 & 0.00 & -0.19 & 0.00 & 0.00 \\ 
  10 &   1 & 21,1 & 20.00 & SBB8 & 6.00 & 0.69 & 0.52 & 0.00 & 0.00 \\ 
  11 &   1 & 1,9 & 1.00 & N & 0.76 & 0.00 & 0.55 & 0.00 & 0.00 \\ 
  12 &   1 & 1,20 & 5.00 & F & 2.82 & 0.00 & 0.29 & 0.00 & 0.00 \\ 
  13 &   1 & 1,2 & 5.00 & F & 13.81 & 0.00 & 0.74 & 0.00 & 0.00 \\ 
  14 &   1 & 2,5 & 5.00 & F & 6.82 & 0.00 & 0.55 & 0.00 & 0.00 \\ 
  15 &   1 & 2,6 & 5.00 & F & 5.97 & 0.00 & 0.51 & 0.00 & 0.00 \\ 
  16 &   1 & 6,4 & 20.00 & SBB8 & 5.90 & 0.53 & 0.38 & 0.00 & 0.00 \\ 
  17 &   1 & 2,3 & 5.00 & F & 4.23 & 0.00 & 0.40 & 0.00 & 0.00 \\ 
  18 &   1 & 9,7 & 5.00 & F & -5.71 & 0.00 & -0.50 & 0.00 & 0.00 \\ 
  19 &   1 & 9,8 & 27.00 & BB1\_90 & -0.10 & -1.73 & -0.45 & 0.00 & 0.00 \\ 
  20 &   1 & 9,10 & 5.00 & F & -4.40 & 0.00 & -0.42 & 0.00 & 0.00 \\ 
   \hline
\end{tabular}
\caption{Summary of the first tree of the vine copula fitted in Cluster 1. utd and ltd correspond to the estimated upper and lower tail dependence of the respective pair copula.}
\end{table}

\begin{table}[ht]
\centering
\begin{tabular}{rrlrlrrrrr}
  \hline
 & tree & edge & family & cop & par & par2 & tau & utd & ltd \\ 
  \hline
1 &   1 & 19,17 & 7.00 & BB1 & 0.56 & 1.11 & 0.30 & 0.14 & 0.33 \\ 
  2 &   1 & 19,15 & 1.00 & N & 0.85 & 0.00 & 0.64 & 0.00 & 0.00 \\ 
  3 &   1 & 17,12 & 17.00 & SBB1 & 0.17 & 2.80 & 0.67 & 0.23 & 0.72 \\ 
  4 &   1 & 17,18 & 17.00 & SBB1 & 0.13 & 1.30 & 0.28 & 0.02 & 0.30 \\ 
  5 &   1 & 18,13 & 17.00 & SBB1 & 0.17 & 2.33 & 0.60 & 0.18 & 0.65 \\ 
  6 &   1 & 15,7 & 5.00 & F & -1.35 & 0.00 & -0.15 & 0.00 & 0.00 \\ 
  7 &   1 & 15,16 & 6.00 & J & 1.35 & 0.00 & 0.17 & 0.33 & 0.00 \\ 
  8 &   1 & 15,11 & 10.00 & BB8 & 2.21 & 0.80 & 0.23 & 0.00 & 0.00 \\ 
  9 &   1 & 13,14 & 10.00 & BB8 & 1.96 & 0.69 & 0.14 & 0.00 & 0.00 \\ 
  10 &   1 & 7,9 & 40.00 & BB8\_270 & -3.31 & -0.74 & -0.34 & 0.00 & 0.00 \\ 
  11 &   1 & 9,21 & 5.00 & F & 3.64 & 0.00 & 0.36 & 0.00 & 0.00 \\ 
  12 &   1 & 9,1 & 2.00 & t & 0.77 & 13.09 & 0.56 & 0.20 & 0.20 \\ 
  13 &   1 & 7,8 & 2.00 & t & 0.48 & 14.16 & 0.32 & 0.04 & 0.04 \\ 
  14 &   1 & 8,10 & 2.00 & t & 0.43 & 17.70 & 0.28 & 0.01 & 0.01 \\ 
  15 &   1 & 1,2 & 1.00 & N & 0.93 & 0.00 & 0.76 & 0.00 & 0.00 \\ 
  16 &   1 & 1,6 & 17.00 & SBB1 & 0.42 & 1.56 & 0.47 & 0.34 & 0.44 \\ 
  17 &   1 & 2,3 & 1.00 & N & 0.61 & 0.00 & 0.42 & 0.00 & 0.00 \\ 
  18 &   1 & 2,20 & 10.00 & BB8 & 5.13 & 0.52 & 0.33 & 0.00 & 0.00 \\ 
  19 &   1 & 2,5 & 17.00 & SBB1 & 0.20 & 1.63 & 0.44 & 0.12 & 0.47 \\ 
  20 &   1 & 6,4 & 1.00 & N & 0.55 & 0.00 & 0.37 & 0.00 & 0.00 \\ 
   \hline
\end{tabular}
\caption{Summary of the first tree of the vine copula fitted in Cluster 2.  utd and ltd correspond to the estimated upper and lower tail dependence of the respective pair copula.}
\end{table}

\clearpage

\raggedright
\bibliography{vineclust_arXiv}

\begin{thebibliography}{22}
\providecommand{\natexlab}[1]{#1}
\providecommand{\url}[1]{\texttt{#1}}
\expandafter\ifx\csname urlstyle\endcsname\relax
  \providecommand{\doi}[1]{doi: #1}\else
  \providecommand{\doi}{doi: \begingroup \urlstyle{rm}\Url}\fi

\bibitem[Aas et~al.(2009)Aas, Czado, Frigessi, and Bakken]{Aas2009}
Kjersti Aas, Claudia Czado, Arnoldo Frigessi, and Henrik Bakken.
\newblock Pair-copula constructions of multiple dependence.
\newblock \emph{Insurance: Mathematics and Economics}, 44\penalty0
  (2):\penalty0 182--198, 2009.

\bibitem[Badih et~al.(2019)Badih, Pierre, and Laurent]{badih2019assessing}
Ghattas Badih, Michel Pierre, and Boyer Laurent.
\newblock Assessing variable importance in clustering: a new method based on
  unsupervised binary decision trees.
\newblock \emph{Computational Statistics}, 34:\penalty0 301--321, 2019.

\bibitem[Bedford and Cooke(2001)]{Bedford2001}
Tim Bedford and Roger~M. Cooke.
\newblock Probability density decomposition for conditionally dependent random
  variables modeled by vines.
\newblock \emph{Annals of Mathematics and Artificial Intelligence},
  32:\penalty0 245–268, 2001.

\bibitem[Bedford and Cooke(2002)]{Bedford2002}
Tim Bedford and Roger~M. Cooke.
\newblock Vines - a new graphical model for dependent random variables.
\newblock \emph{Annals of Statistics}, 30\penalty0 (4):\penalty0 1031--1068,
  2002.

\bibitem[Breiman(2001)]{breiman2001random}
Leo Breiman.
\newblock Random forests.
\newblock \emph{Machine Learning}, 45:\penalty0 5--32, 2001.

\bibitem[Clelland and Hill(2019)]{Clelland2019}
David Clelland and Carol Hill.
\newblock {Deprivation, policy and rurality: The limitations and applications
  of area-based deprivation indices in Scotland}.
\newblock \emph{Local Economy: The Journal of the Local Economy Policy Unit},
  34\penalty0 (1):\penalty0 33–50, 2019.
\newblock \doi{10.1177/0269094219827893}.

\bibitem[Executive(2005)]{ScotExecutive2005}
Scottish Executive.
\newblock Scottish neighbourhood statistics: Intermediate geography background
  information.
\newblock Technical report, Scotland: Scottish Executive, 2005.

\bibitem[Funk et~al.(2024)Funk, Ludwig, Kuechenhoff, and
  Nagler]{funk2024towards}
Henri Funk, Ralf Ludwig, Helmut Kuechenhoff, and Thomas Nagler.
\newblock {Towards more realistic climate model outputs: A multivariate bias
  correction based on zero-inflated vine copulas}.
\newblock \emph{arXiv preprint arXiv:2410.15931}, 2024.

\bibitem[Hartigan and Wong(1979)]{Hartigan1979}
J.~A. Hartigan and M.~A. Wong.
\newblock {Algorithm AS 136: a k-means clustering algorithm}.
\newblock \emph{Journal of the Royal Statistical Society. Series C (Applied
  Statistics)}, 28\penalty0 (1):\penalty0 100--108, 1979.
\newblock \doi{10.2307/2346830}.

\bibitem[He et~al.(2005)He, Cai, and Niyogi]{he2005laplacian}
Xiaofei He, Deng Cai, and Partha Niyogi.
\newblock Laplacian score for feature selection.
\newblock \emph{Advances in Neural Information Processing Systems}, 18, 2005.

\bibitem[Li and Joe(2024)]{li2024multivariate}
Xiaoting Li and Harry Joe.
\newblock Multivariate directional tail-weighted dependence measures.
\newblock \emph{Journal of Multivariate Analysis}, 203:\penalty0 105319, 2024.

\bibitem[McCartney and Hoggett(2023)]{McCartney2023}
G.~McCartney and R.~Hoggett.
\newblock How well does the scottish index of multiple deprivation identify
  income and employment deprived individuals across the urban-rural spectrum
  and between local authorities?
\newblock \emph{Public Health}, 217:\penalty0 26–32, 2023.
\newblock \doi{10.1016/j.puhe.2023.01.009}.

\bibitem[McKendrick et~al.(2011)McKendrick, Barclay, Carr, Clark, Holles,
  Perring, and Stein]{McKendrick2011}
JH~McKendrick, C~Barclay, C~Carr, A~Clark, J~Holles, E~Perring, and L~Stein.
\newblock \emph{Our rural numbers are not enough: an independent position
  statement and recommendations to improve the identification of poverty,
  income inequality and deprivation in rural Scotland}.
\newblock Rural Poverty Indicators Action Learning Set, March 2011.

\bibitem[Meng and Rubin(1993)]{Meng1993}
Xiao-Li Meng and Donald~B. Rubin.
\newblock {Maximum likelihood estimation via the ECM algorithm: a general
  framework}.
\newblock \emph{Biometrika}, 80\penalty0 (2):\penalty0 267--278, 1993.
\newblock \doi{10.2307/2337198}.

\bibitem[Panagiotelis et~al.(2012)Panagiotelis, Czado, and
  Joe]{panagiotelis2012pair}
Anastasios Panagiotelis, Claudia Czado, and Harry Joe.
\newblock Pair copula constructions for multivariate discrete data.
\newblock \emph{Journal of the American Statistical Association}, 107\penalty0
  (499):\penalty0 1063--1072, 2012.

\bibitem[Sahin(2021)]{vineclust}
{\"O}zge Sahin.
\newblock \emph{vineclust}, 2021.
\newblock URL \url{https://github.com/oezgesahin/vineclust}.

\bibitem[Sahin and Czado(2022)]{sahin2022}
{\"O}zge Sahin and Claudia Czado.
\newblock {Vine copula mixture models and clustering for non-Gaussian data}.
\newblock \emph{Econometrics and Statistics}, 22:\penalty0 136--158, 2022.
\newblock \doi{10.1016/j.ecosta.2021.08.011}.

\bibitem[Schwarz(1978)]{Schwarz1978}
Gideon Schwarz.
\newblock Estimating the dimension of a model.
\newblock \emph{The Annals of Statistics}, 6\penalty0 (2):\penalty0 461--464,
  1978.
\newblock \doi{10.1214/aos/1176344136}.

\bibitem[Scrucca et~al.(2016)Scrucca, Fop, Murphy, and Raftery]{Scrucca2016}
Luca Scrucca, Michael Fop, T.~Brendan Murphy, and Adrian~E. Raftery.
\newblock {Mclust 5: Clustering, classification and density estimation using
  Gaussian finite mixture models}.
\newblock \emph{R Journal}, 8\penalty0 (1):\penalty0 289--317, 2016.
\newblock \doi{10.32614/rj-2016-021}.

\bibitem[Senior(2019)]{Senior2019}
Steven~L Senior.
\newblock Using hierarchical clustering to explore patterns of deprivation
  among english local authorities.
\newblock \emph{Journal of Public Health}, 42\penalty0 (4):\penalty0 772–777,
  December 2019.
\newblock ISSN 1741-3850.
\newblock \doi{10.1093/pubmed/fdz182}.
\newblock URL \url{http://dx.doi.org/10.1093/pubmed/fdz182}.

\bibitem[Shi and Horvath(2006)]{shi2006unsupervised}
Tao Shi and Steve Horvath.
\newblock Unsupervised learning with random forest predictors.
\newblock \emph{{Journal of Computational and Graphical Statistics}},
  15\penalty0 (1):\penalty0 118--138, 2006.

\bibitem[SIMD(2020)]{SIMD2020}
SIMD.
\newblock Scottish index of multiple deprivation.
\newblock Technical report, Scottish Government, 2020.

\end{thebibliography}
\end{document}